\documentclass[12pt]{article}
\usepackage{caption}
\usepackage{amssymb,amsmath,mathtools,graphicx,xspace,colortbl,ragged2e,rotating}
\usepackage{textcomp}
\usepackage{amsfonts}
\usepackage{indentfirst}
\usepackage{microtype}
\usepackage{amsmath}
\usepackage{yfonts}
\usepackage{bbm}
\usepackage{dsfont}
\usepackage{braket}
\usepackage{dsfont}
\usepackage{subcaption}
\usepackage{caption}
\usepackage{tikz}
\usepackage{xcolor}
\usepackage{float}
\usepackage{verbatim}
\usepackage{hyperref}
\usepackage{xcolor}
\usepackage{tocloft}
\usepackage{amsmath,amsfonts,amssymb}
\usepackage{mathtools}
\usepackage{hyperref}
\usepackage{placeins}
\usepackage[bbgreekl]{mathbbol}
\usepackage{graphicx}
\usepackage{color}
\usepackage{epsfig}
\usepackage{psfrag}	
\usepackage{tikz}
\usetikzlibrary{snakes}
\usepackage{slashed}
\usepackage{ulem}
\usepackage[font=small,labelfont=bf,width=1\textwidth]{caption}
\usepackage{comment}
\usepackage[all]{xy}
\usepackage{multirow}
\usepackage{float}
\usepackage{tensor}
\usepackage {multirow}
\usepackage {booktabs}
\usepackage {fancyhdr}
\usepackage[T1]{fontenc}

\numberwithin{equation}{section}
\DeclareMathOperator{\Tr}{Tr}

\newcommand{\bea}{\begin{eqnarray}}
\newcommand{\eea}{\end{eqnarray}}
\newcommand{\beq}{\begin{equation}}
\newcommand{\eeq}{\end{equation}}
\newcommand{\bal}{\begin{equation}\begin{aligned}}
\newcommand{\eal}{\end{aligned} \end{equation}}

\newcommand{\address}[1]{\vbox{\center\em#1}}
\renewcommand{\title}[1]{\vbox{\center\huge{#1}}\vspace{5mm}}

\usetikzlibrary{decorations.markings}
\graphicspath{{../graphics/}{../tcache/}{../gcache/}}
\DeclareGraphicsExtensions{.pdf,.eps,.ps,.png,.jpg,.jpeg}

\begin{document}
\begin{titlepage}
\begin{center}
\vspace*{20mm}

\title{Nielsen complexity with multiple cost factors}
\vspace{5mm}

Marcos Rios Ribeiro$^a$ and Diego Trancanelli$^b$
\vskip 3mm
\renewcommand{\thefootnote}{$\alph{footnote}$}

\address{$^a$Department of Mathematical Physics, Institute of Physics \\ University of S\~ao Paulo,
05314-970 S\~ao Paulo, Brazil}

\address{$^b$Dipartimento di Scienze Fisiche, Informatiche e Matematiche, \\
Universit\`a di Modena e Reggio Emilia, via Campi 213/A, 41125 Modena, Italy \\ \& \\
INFN Sezione di Bologna, via Irnerio 46, 40126 Bologna, Italy}
\vskip 5mm

\tt{marcosribeiro@usp.br, diego.trancanelli@unimore.it}

\renewcommand{\thefootnote}{\arabic{footnote}}
\setcounter{footnote}{0}
\end{center}

\vspace{8mm}
\abstract{
\normalsize{
\noindent
We investigate Nielsen’s geometric approach to quantum complexity in the presence of multiple cost factors, extending the standard framework where a single penalty distinguishes easy from hard directions of the group manifold. By introducing a hierarchy of penalties associated with different degrees of non-locality, we develop a generalized right-invariant complexity geometry and analyze its implications for geodesic evolution. We derive the modified Euler–Arnold and Jacobi equations and study how multiple cost factors reshape the structure and scaling of conjugate points, where geodesic optimality breaks down. The formalism is illustrated in two settings: a single-qubit system with two cost factors, where we develop a semi-analytic approach to investigate complexity growth and its dependence on penalty hierarchies, and SYK-type models, where we analyze both free and chaotic regimes. In these many-body systems, we show that distinct non-local sectors generate multiple families of conjugate points whose occurrence depends on both the cost hierarchy and the system size. Our results highlight how refining the penalty structure provides a richer and more realistic description of quantum complexity and its dynamical behavior.}}
\vfill
\end{titlepage}
\tableofcontents


\section{Introduction}

Quantifying the `difficulty' of performing a given task is one of the central goals of computer science, leading to the proposal of numerous notions of {\it computational complexity} \cite{papadimitriou2003computational}. In recent years, this goal has found its way into quantum physics, since quantum evolution itself can be thought as a computational task, where an initial quantum state or operator are transformed into given targets, see \cite{Baiguera_2026} for a recent review. As a result, ideas based on complexity have been applied across several areas of physics, from many-body systems - where complexity has been proposed as a probe for phases of matter, including topological ones \cite{bhattacharyya2018circuit,Ali_2020,Balasubramanian_2020,balasubramanian2021complexity,Craps:2022ese,alhambra2023quantum,Caputa_2022,Craps:2023rur}, -  to high-energy physics - where it is conjectured to play an essential role in quantum gravity and black hole physics \cite{Balasubramanian_2011,susskind2014entanglement,Brown_2016,jefferson2017circuit}.

Despite the explosion of work on quantum complexity in recent years, there is still no consensus about a  precise definition of quantum complexity, see for example \cite{parker2019universal,camargo2019path}. In fact, depending on the physical setting, it can be more convenient to work with one notion of complexity rather than another. This lack of a single definition has become a recurring subject in the recent literature \cite{grimm2025complexityquantumfieldtheory}. 

\begin{figure}[H]
    \centering
    \includegraphics[scale=0.4]{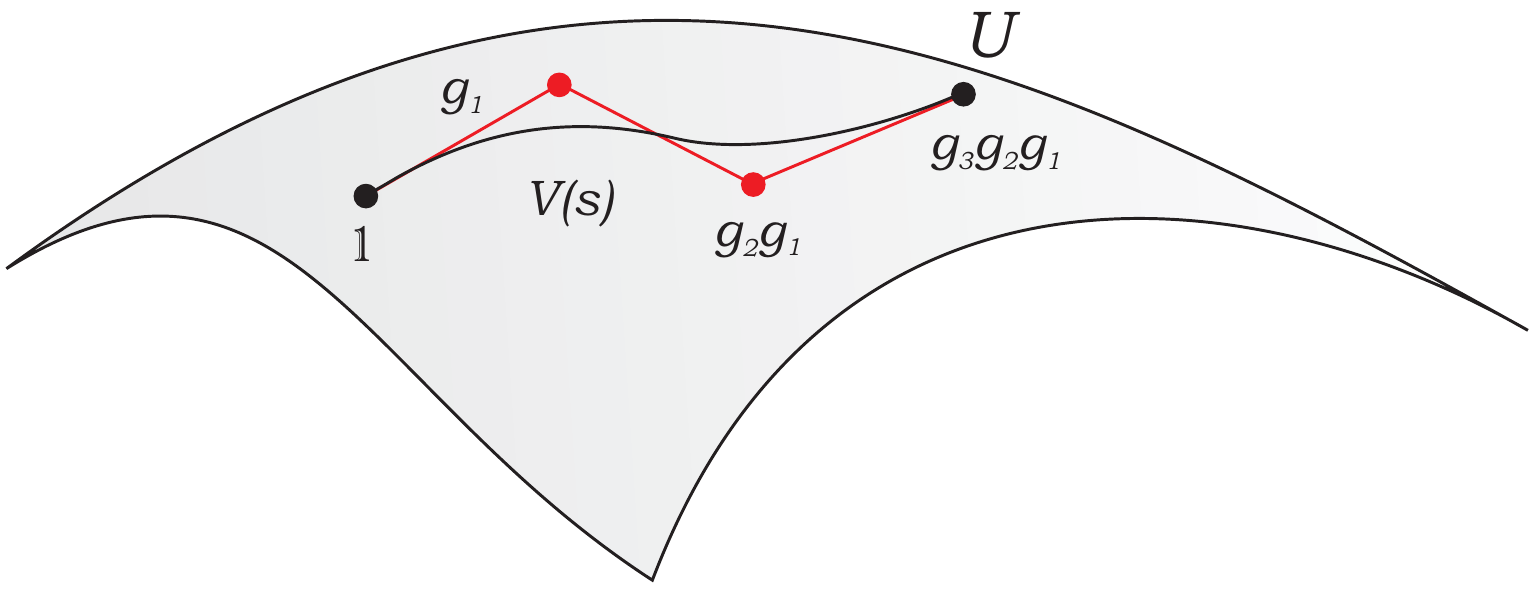}
    \caption{Schematic representation of a unitary manifold and a quantum circuit as a sequence of elementary gates. A geodesic path (black) is depicted from the identity to some target unitary $U$. The red straight segments represent the construction of a circuit using some elementary gates $g_i$. In this example, the final unitary is $U = g_3g_2g_1$. The geodesic is parametrized by a parameter $s$ and it approximates the circuit smoothly by varying a control velocity $V (s)$, analogous to an infinitesimal elementary gate.}
    \label{figure1}
\end{figure}

Among the various proposals, Nielsen's geometric approach \cite{nielsen2005geometric,Nielsen_2006,dowling2008geometry} is particularly interesting, as it provides a very concrete formulation of complexity in terms of geodesics on the group manifold associated with the quantum system at hand. In this framework, the complexity of a given quantum operation is defined as the length of the shortest path connecting an initial and a target unitary, seen as two points of the group manifold. This length is measured with a right-invariant Finsler metric that assigns different costs to different directions in the Lie algebra, favoring propagation along so-called {\it easy (or local) directions} of the manifold compared to the {\it hard (or non-local) directions}. A key advantage of Nielsen's formalism, often referred to as {\it complexity geometry}, is that it can be formulated in both discrete and continuous settings \cite{Brown_2019,chapman2018toward}, see figure~\ref{figure1}.

The geometric features of the manifold are directly related to the physical content of complexity. For instance, negative curvature is often regarded as desirable \cite{brown2017quantum,Brown_2019}. At the same time, positive curvature is inevitable in some scenarios \cite{Balasubramanian_2020,Auzzi_2021} and it can lead to geodesic focusing phenomena and the occurrence of so-called {\it conjugate points}. It is crucial to study these, because a geodesic fails to be locally minimizing after crossing its first conjugate point, thus ceasing to represent the true complexity. Consequently, understanding the location and scaling of conjugate points is essential for describing the growth of complexity as a geodesic length, including possible obstructions. In previous works \cite{Balasubramanian_2020,balasubramanian2021complexity}, a metric that separates easy and hard directions through a large cost factor $\mu$ was adopted to study the emergence and location of conjugate points associated with these directions. In particular, conjugate points associated with hard directions typically scale as $t_c \sim \mu$, while in chaotic systems local obstructions are expected to be absent until exponential times.

In this work, we investigate Nielsen's complexity geometry in the presence of multiple cost factors, \emph{i.e.} a family of right-invariant complexity metrics in which distinct classes of hard directions are assigned different penalties $\mu_1 <\mu_2<\dots$. Although ideas of penalty structures have been explored in other contexts \cite{Auzzi_2021,Brown_2023,Chowdhury:2023iwg,Bhattacharyya:2024rzz,brown2024polynomial,baiguera2025cftcomplexitypenaltyfactors,PhysRevLett.132.160402, PhysRevLett.134.050402,Craps:2025kub}, our focus here is on how the internal structure among hard directions reshapes both the geodesic dynamics and the conjugate points. We generalize the analysis of the Euler-Arnold and Jacobi equations to multiple penalties and provide both analytical and numerical checks of the resulting effects. We then illustrate the framework in two examples. First, we study a one-qubit model with two cost factors, where the Euler-Arnold equations can be solved exactly, while the corresponding Dyson equation is treated numerically to reconstruct the geodesics and determine the complexity. Then we analyze the emergence and scaling of local and non-local conjugate points in SYK-type models, and show how their first occurrences depend on the hierarchy of penalties and on the growth of non-local directions with system size. 

The paper is organized as follows. In section \ref{sec:theory}, after a brief review of Nielsen's geometric complexity with a single cost factor, we analyze how the Euler-Arnold equations and the conjugate points get modified by the introduction of multiple penalties. In section \ref{sec:1qubit}, we apply the formalism to a single qubit, which we study in detail introducing two different cost factors and considering three sample Hamiltonians. In section \ref{sec:SYK}, we study SYK models, with both quadratic (or {\it free}) and cubic and quartic (or {\it chaotic}) Hamiltonians, again in the presence of two different cost factors. Finally, we conclude with some outlook, while a few cumbersome expressions needed for the analysis in section \ref{sec:SYK} are relegated to an appendix.


\section{Complexity with multiple cost factors}
\label{sec:theory}

\subsection{Review of the single cost factor case}

To put things into context and to set up notation, we start with a brief review of Nielsen's geometric complexity in the usual case in which a single cost factor is contemplated. We follow closely \cite{Nielsen_2006}.

Let us consider a system of $N$ qubits and its Hilbert space $ \mathcal{H}=(\mathbb{C}^{2})^{\otimes N}$. The relevant symmetry group is $SU(2^N)$, with generators given by tensor products of the Pauli matrices and the identity, the so-called {\it generalized Pauli matrices}:
\begin{equation}
T_{I}=\sigma _{I_{1}}\otimes \sigma _{I_{2}}\otimes \ldots \otimes \sigma
_{I_{N}}.
\end{equation}%
Here $I$ is a collective index $I_1 I_2 \cdots I_N$, where for any qubit we have $\sigma_{I=0,1,2,3}=(\mathbbm{1},\vec \sigma)$. These matrices are hermitian and, apart from the identity $T_0=\mathbbm{1}^{\otimes N}$, traceless. They satisfy the algebra 
\begin{equation}
\lbrack T_{I},T_{J}]= i f_{IJ}^{K}T_{K},
\end{equation}%
with structure constants $f_{IJ}^{K}$.

Each point of the $SU(2^N)$ group manifold is associated with a unitary operator $U$ and the relative complexity of two different operators $U$ and $\tilde U$, denoted as $\mathcal{C} (U,\tilde U)$, can be interpreted as the distance in $SU(2^N)$ between their corresponding points. To understand what is precisely meant by `distance', one first has to define a path in $SU(2^N)$ between (the points corresponding to) $U$ and $\tilde U$ as a family of unitary operators $V(s)$ parameterized by $s \in [0,1]$ and governed by the Dyson equation
\begin{equation}
\frac{dV(s)}{ds}=- i h(s)V(s),  \label{Dysonequation}
\end{equation}%
with initial and final conditions 
\begin{equation}
V(0)=U,\qquad V(1)=\tilde U.
\end{equation}%
This depends on the local Hamiltonian $h(s)$, which should be a linear combination of the Lie algebra elements, thus admitting a decomposition in terms of the generalized Pauli matrices 
\begin{equation}
h(s)=\sum_{I}\gamma ^{I}(s)\, T_{I},
\end{equation}%
where the $\gamma ^{I}(s)$ are called \emph{control functions}, collected in a vector $\gamma(s)$ with $4^{N}-1$ components. 

Finding the `optimal' values of the control functions is equivalent to finding the relative complexity $\mathcal{C}(U,\tilde U)$. This optimality is achieved by first defining a \emph{cost function} associated with the control functions 
\begin{equation}
c_{f}({\gamma})\equiv \int_{0}^{1}ds\, f\left( \gamma (s)\right) , \label{gamma-cost}
\end{equation}%
where $f:\mathbb{R}^{4^{N}-1}\rightarrow \mathbb{R}$ is a functional mapping the space in which $\gamma(s)$ lives into the real set. We also define the relative cost associated with the reference operator $U$ and the target operator $\tilde U$ as 
\begin{equation}
c_{f}(U,\tilde U)\equiv \min_{\gamma}c_{f}(\gamma)
\label{U-cost}.
\end{equation}%
We would like $c_{f}({\gamma})$ to capture the notion of complexity associated with ${\gamma}(s)$, and therefore with $h(s)$, and $c_{f}(U,\tilde U)$ to be related to the complexity $\mathcal{C} (U,\tilde U)$.

Notice that all these definitions depend on the function $f$, that needs to obey certain properties in order for (\ref{gamma-cost}) and (\ref{U-cost}) to be good definitions of measure:
\begin{itemize}
    \item \emph{Continuity} is required for obvious reasons, since we do not expect a small perturbation on the control function to produce a large change in its cost.
    \item \emph{Homogeneity} is also a good requirement because, when one rescales $h(s)$, one would like the cost to behave in the same way, {\it i.e.} $f(a \gamma)=af(\gamma)$ for any $\gamma$.
    \item \emph{Positivity} is also required, then $f(\gamma) \geq 0$ with the equality holding if and only if $\gamma=0$.
    \item  \emph{Triangle inequality} is a crucial property for a good measure. Then the cost function must satisfy $f(x+y)\leq f(x)+f(y)$.    
\end{itemize}

The cost functions can be now reformulated in geometric terms, which is something very common in Hamiltonian control problems \cite{todorov2006optimal}. First of all, let us introduce the notion of metric on $SU(2^N)$ as a map on the tangent bundle $T[SU(2^N)]$, represented by the pair $(x,y)$, where $x$  is an element of the group manifold and $y$ belongs to the corresponding tangent space. Then $F: T[SU(2^N)] \to \mathbb{R}^+$ is defined such that, for any $x \in SU(2^N)$, the function is positive definite $F(x, y) \geq 0$, homogeneous in the second variable $F(x, ay) = a F(x, y)$ for $a \geq 0$, and satisfies the triangle inequality $F(x, y + z) \leq F(x, y) + F(x, z)$. This metric function is known as the \emph{Finsler metric} and it generalizes the Riemannian metric, as it accounts for other types of functional measure beyond the usual quadratic form.

Given a metric $F$, we can introduce the notion of the length of a curve $u(s): I \subseteq \mathbb{R} \to SU (2^N) $ as follows
\begin{equation}
\ell_F(u)\equiv \int_I ds F(u(s),\dot{u}(s)), \label{curvelength}   
\end{equation}
where $\dot{u}(s)$ is a tangent vector to $u(s)$. It is easy to show that the length is invariant under reparameterizations and we can then always take $I=[0,1]$. The distance $d(U,\tilde U)$ between two points $U$ and $\tilde U \in SU(2^N)$ is defined as the minimum length between them
\begin{equation}
    d(U,\tilde U)=\min_{u}\int^{1}_0 ds F(u(s),\dot{u}(s)),
\end{equation}
where the minimum is evaluated over all curves $u(s)$ subject to
\begin{align}
    u(0)=U,\qquad u(1)=\tilde U.
\end{align}

Finally, we motivate the connection between the cost function and the Finsler metric. We can define a metric in $SU(2^N)$ by $F(x,y)\equiv f(\gamma )$, where $f$ is a cost function and $\gamma$ is the vector control function. This is a good definition, since $f$ satisfies all required conditions for a Finsler metric. Moreover, this identification of tangent vectors with control functions ensures that the coordinates on the tangent space are just the control functions $\gamma^I$. Therefore, it justifies that the complexity defined by \eqref{U-cost} be expressed in geometric terms through the distance between operators
\begin{equation}
    \mathcal{C}(U,\tilde U)\equiv \min_V\int^{1}_0 ds\,  F\left(V(s),h(s)\right).
\end{equation}

The metric discussed so far is a special type of metric, called the \emph{right-invariant metric}, meaning that it is invariant under right-multiplication
\begin{equation}
\label{rightinv}
    F\left(x,y\right)=F\left(xz,r_z(y)\right),
\end{equation}
where $r_z(y)$ is a right-multiplication pushforward that connects the tangent spaces $T_{x}[SU(2^N)]$ and $T_{xz}[SU(2^N)]$. The right-invariant property naturally extends to complexity, thus it follows that
\begin{align}
    \mathcal{C}(U,\tilde U)=\mathcal{C}(U W,\tilde U W), \quad\quad \forall\,  W\in SU(2^N).
\end{align}
Due to this property, the distance between two points $U$ and $\tilde U$ can be reduced to the distance from the identity to the point $\tilde U U^{-1}$.

It is a standard result in the calculus of variations \cite{arnol2013mathematical} that any curve $V(s)$ which is an extreme of the functional $\ell_F(u)$ must satisfy the Euler-Lagrange equations. However, it is convenient for us to work with a different form of those equations, called the {\it Euler-Arnold equations} \cite{Balasubramanian_2020}. In a coordinate basis, these take the form
\begin{equation}
    G_{IJ}\frac{dy^{J}}{ds}= f_{IJ}^K\,y^{J}G_{KL}y^{L},
    \label{geodesics}
\end{equation}
where $y^{J}(s)$ are the components of the tangent vector to the curve $V(s)$, defined by expanding $i\dot{V}(s)V^{\dagger}(s)$ in a fixed basis as $i\dot{V}(s)V^{\dagger}(s)=\sum_J y^{J}(s)T_J$, $G_{IJ}$ is the Finsler metric tensor and $f_{IJ}^{K}$ are the structure constants of the Lie algebra associated with the group manifold. The Finsler metric tensor is related to the metric function $F$ through
\begin{equation}
    G_{IJ}=\frac{1}{2}\frac{\partial^2F^{2}}{\partial y^I \partial y^J}\label{Finslertensormetric}.
\end{equation}

The important point here is that $G_{IJ}$ should describe an anisotropic space. The reason for this is quite simple: some directions in the tangent space $T_U[SU(2^N)]$ are easier to follow than others. In general, we can assume that the directions associated with fewer Pauli matrices are simpler than those involving many Pauli matrices. In fact, this reflects an arbitrariness that arises in various ways in studies of complexity, when trying to classify what is considered easy and what is considered hard. In real physical systems, a pragmatic approach is to identify the largest number of Pauli matrices appearing in a given Hamiltonian, say $k$, and to classify operators or directions with a number of Pauli matrices $n\leq k$ as {\it easy} or {\it local} directions, while those with $n>k$ as {\it hard} or {\it non-local} directions. This is called $k${\it-locality}. To encode this difference between easy and hard direction we work with the metric
\begin{equation}
    G_{IJ}=\Omega_{I}\delta_{IJ}\label{tensormetric},
\end{equation}
where there is no summation on the repeated indices on the right-hand side and $\Omega_I$ is given by
\begin{equation}\label{pd}
    \Omega_I =
\begin{cases}
1 & \text{for $I \in $ easy directions } \\
1 +\mu & \text{for $I \in $ hard directions } 
\end{cases}.
\end{equation}
In the following we are going to denote indices corresponding to easy direction with undotted Greek letters, $\alpha,\beta,\ldots$, while indices corresponding to hard directions are going to be denoted by dotted Greek indices, $\dot\alpha,\dot\beta,\ldots$. The constant $\mu$ is called the \emph{cost} or {\it penalty factor} because it discourages propagation in non-local directions. This respects the right-invariant property of the Finsler metric \eqref{rightinv} and when $\mu = 0$ we recover the so-called  \emph{bi-invariant} case. 

The geodesic equation (\ref{geodesics}) produces a path on $SU(2^N)$ that follows the Dyson equation (\ref{Dysonequation}) and whose formal solution is given by the path ordered exponential
\begin{equation}
    V(s)= \mathcal{P}\exp\left(-i\int_{0}^{s}ds'\,  y^{I}(s')T_I\right).
\end{equation}
The relative complexity of its boundary points $V(0)=U$ and $V(1)=\tilde U$ is given by the length of the geodesics connecting them, namely
\begin{equation}
    \mathcal{C}(U,\tilde U)=\int_{0}^1 ds\sqrt{G_{IJ}y^I(s)y^J(s)}
    \label{relativecomplexity}.
\end{equation}

Next, we move on to reviewing conjugate points, following \cite{Balasubramanian_2020,balasubramanian2021complexity}. These are essentially intersection points of different geodesics and characterize the loss of local minimality of a geodesic. Although their existence does not by itself determine the loss of global minimality, they provide an important local diagnostic of geodesic optimality. The loss of global minimality is instead associated with the so-called {\it cut locus}, which can in general be reached before the first conjugate point, as stressed in \cite{susskind2014switchbacksbridge}.

The idea is to start with a geodesic $V:[0,1]\to SU(2^{N})$ with endpoints\footnote{We can always reparametrize the interval to get the endpoint at $s=1$.} $V(0)=U$ and $V(1)=\tilde U$, together with a family of $\epsilon$-deformed geodesics, $V_{\epsilon}:[0,1]\to SU(2^N)$, defined as $V_{\epsilon}(s)=V(s)+\epsilon \, \delta V(s)+\mathcal{O}(\epsilon^2)$, where the variation $\delta V(s)$ is known as a {\it Jacobi field}. The $\epsilon$-deformed geodesics are generated by an $\epsilon$-deviation in the velocities, $y^{I} \to y^{I}+\epsilon\, \delta y^{I}$. One can find $\delta y^{I}$ by solving the linearized Euler-Arnold equations which, in this context, are called {\it Jacobi equations}. We say that $U$ is conjugate to $\tilde U$ with respect to the geodesic $V$ if there exists a nontrivial Jacobi field that satisfies
\begin{equation}
\delta V(0)=0, \quad \quad \delta V(1)=\mathcal{O}(\epsilon^2).\label{reqconjugates}
\end{equation}

We then need to check if there exist conjugate points on a geodesic between two points
on $SU(2^N)$, say an origin $U$ and a target operator $\tilde U$. If the answer is yes,
then beyond the first conjugate point this geodesic ceases to be locally minimizing.
To find the location of conjugate points, we need to translate the conditions \eqref{reqconjugates}
in terms of velocities.

Consider a generic geodesic $V(s)$ generated by the local Hamiltonian $h(s)=y^{I}(s)T_{I}$. A small deviation from it produces an $\epsilon$-deformed geodesic that satisfies the following Dyson equation
\begin{equation}
\frac{d V_{\epsilon}}{ds}=-i\left(h+\epsilon\, \delta y\right)V_{\epsilon},
\end{equation}
where we omitted the $s$-dependence. The key point here is that we can obtain a simple equation for the Jacobi field at first order in $\epsilon$. It reads
\begin{equation}
\frac{d \delta V}{ds}=-i h\, \delta V-i\,  \delta y V +\mathcal{O}(\epsilon^2),
\end{equation}
which can be solved easily, by multiplying on the left by an integrating factor $F(s)$, to get
\begin{equation}
\delta V(s)=-iF^{-1}(s)\int^{s}_{0}ds'F(s')\delta y(s')V(s').\label{jacobiF}
\end{equation}
Here $F(s)$ satisfies the equation $\dot F=iF h$, which has a formal solution in terms of the anti-path-ordered exponential\footnote{This is nothing but the Hermitian conjugate of the standard path-ordered exponential.}
\begin{equation}
F(s)=\mathcal{\bar P}\exp\left (i\int_{0}^{s}ds'h(s')\right).
\end{equation}
Note that our expression for the Jacobi fields \eqref{jacobiF} already satisfies the first requirement in \eqref{reqconjugates}. The second requirement follows if
\begin{equation}
    \int^{1}_{0}ds\, F(s)\delta y(s)V(s)=0.
\end{equation}
Finding conjugate points is mapped into finding deformed velocities $\delta y(s)$ that satisfy the constraint above and are specified by an initial condition $\delta y(0)$ (since the equations are first order). This motivates to consider a {\it Jacobi (super)operator} $\mathbf{Y}_{\mu}:T[SU(2^{N})]\to T[SU(2^{N})]$, defined by
\begin{equation}
    \mathbf{Y}_{\mu}[\delta y(0)]\equiv\int^{1}_{0}dsF(s)\delta y(s)V(s),
\end{equation}
with the conjugate points being the zero-modes of this operator
\begin{equation}
    \delta V(1)=\mathcal{O}(\epsilon^2)\quad \Longleftrightarrow \quad \mathbf{Y}_{\mu}[\delta y(0)]=0 .
\end{equation}

In principle, this formalism can be applied to any geodesic. However, solving the Jacobi equations can be a challenging task. For this reason, in this work we focus on exponential geodesics, which are generated by constant solutions. Moreover, let us consider a local solution, which means that the only non-zero components are the easy/local ones
\begin{equation}
    y^{I}=\delta ^{I}_{\alpha} v^{\alpha} \quad \implies \quad V(s)=e^{-i v s}, \qquad v=v^\alpha T_\alpha.
\end{equation}
The separation of easy/hard directions leads to two different sectors in the linearization procedure of the Jacobi equations:
\begin{equation}
      \frac{d \delta y^{\alpha}}{ds}=\mu f_{\beta \dot\alpha}{}^{\alpha}v^{\beta}\delta y^{\dot\alpha}, \qquad
      \frac{d \delta y^{\dot\alpha}}{ds}=\frac{\mu}{1+\mu} f_{\alpha\dot\beta}{}^{\dot\alpha}v^{\alpha}\delta y^{\dot\beta}. \label{jacobieqs}
\end{equation}

As an illustration of how to find the location of conjugate points using this formalism, we consider the bi-invariant case $\mu= 0$. In this limit, the solutions are simply constants $\delta y(s)=\delta y(0)$ and the conjugate points are determined by the following constraint 
\begin{equation}
   \mathbf{Y}_{0}[\delta y(0)]=  \int^1_0 ds \, e^{ i v s}\delta y(0)e^{- i vs}=0\label{Ylocal}.
\end{equation}
We can choose the initial velocity as the outer product of eigenstates $\{\ket{\tau_i}\}$ of the operator $v$, that is, $\delta y(0)=\ket{\tau_i}\bra{\tau_j}$. This means that $\ket{\tau_i}\bra{\tau_j}$ is an eigenstate of the Jacobi operator $\mathbf{Y}_{0}$, leading to a straightforward integration of \eqref{Ylocal}
\begin{equation}
\mathbf{Y}_{0}\left[\ket{\tau_i}\bra{\tau_j}\right]=\phi(\Delta\tau_{ij})\ket{\tau_i}\bra{\tau_j},
\end{equation}
where $\Delta\tau_{ij}=\tau_i-\tau_j$ and $\phi(x)=(e^{ i x}-1)/ix$. The zero-modes of the Jacobi operator occur whenever the eigenvalue is null, namely $\Delta\tau_{ij}=2\pi$, otherwise the conjugate points associated with that specific direction never occur. 

To recast these results in more physical terms, we can think of the following identifications
\begin{align}
\text{geodesic } V(s) \quad \longleftrightarrow\quad  &\text{unitary time evolution } U(t), \nonumber
\\
\text{generator } v \quad \longleftrightarrow\quad  &\text{physical Hamiltonian } H ,
\label{dictionary}
\end{align}
where by $H$ we mean the actual Hamiltonian of the system considered. For constant solutions of the Euler-Arnold equations, this is implemented in practice as
\begin{equation}
\label{dictionarydefs}
v \mapsto tH, \qquad \tau_i \mapsto E_i t,    
\end{equation}
so that
$
\Delta \tau_{ij} \mapsto \Delta E_{ij}\, t.$
This time-rescaling introduces the physical notion of conjugate time for $E_{i}\ne E_{j}$,
\begin{equation}
    t_{c}=\frac{2\pi n}{\Delta E_{ij}}, \qquad n\in \mathbb{Z}.
\end{equation}
In this context, the quantum time evolution $e^{-iHt}$ is no longer the least complex evolution from $t_c$ onward.

When the cost factor is non-zero, we get a more complicated scenario, which we discuss in the next section, generalizing it at the same time to multiple cost factors.


\subsection{Introducing multiple cost factors}

Typically, in an $N$-qubit system the number of non-local generators is much larger than the number of local ones. For a specific choice of $k$-locality, the number of easy directions scales in fact like $N^k$, while the number of hard ones scales like $4^N$. Moreover, there is a sense in which not all local directions should be treated equally, and the same applies to the non-local ones. For example, one could want to distinguish between easy directions with $k=2$ in which a gate acts on neighboring sites from the ones in which the two sites are far away from each other \cite{deBuruaga:2025val}. With this distinction, based on the extra requirement of adjacency of the sites on which an easy gate acts, the number of easy directions becomes linear in $N$. These `not-so-easy'  directions corresponding to non-neighboring sites become then part of the non-local directions. Similarly, it should be easier to act on three gates at the same time than to act on, say, 9 gates, so that there should also be a hierarchy of hard directions. 

These considerations motivate the introduction of multiple degrees of non-locality and multiple associated cost factors. A geometry that captures these different notions of non-locality is a simple generalization of the right-invariant metric (\ref{tensormetric}), this time with the proportionality factor $\Omega_I$
\begin{equation}\label{omega}
    \Omega_I =
\begin{cases}
1 & \text{for $I \in $ easy directions } \\
1 +\mu_p & \text{for $I \in $ $p$-hard directions } 
\end{cases},
\end{equation}
where the cost factors $\mu_p$ can be taken to be different. This allows to parameterize different degrees of non-locality, by appropriately tuning the values of these cost factors.

We analyze now the implications of this generalization, starting from the geodesics. The right-invariant metric with (\ref{omega}) reads explicitly 
\begin{equation}
G_{IJ}=\left( 
\begin{array}{cccc}
\delta_{\alpha\beta} & 0 & \cdots & 0 \\ 
0 & \left( 1+\mu_{1}\right) \delta_{\dot{\alpha}_{1}\dot{\beta}_{1}} & \cdots
& 0 \\ 
\vdots & \vdots & \ddots & \vdots \\ 
0 & 0 & \cdots & \left( 1+\mu_{D}\right) \delta_{\dot{\alpha}_{D}\dot{\beta }%
_{D}}%
\end{array}
\right).  
\label{full-metric}
\end{equation}
As above, undotted Greek indices represent local directions, while dotted ones the non-local ones. In addition, the non-local directions have now another index $p=1,\ldots , D$ representing the degree of non-locality ($D$ is the number of non-local subspaces). Therefore, $T_{\dot{\alpha}_p}\in \mathrm{NL}_p$ represents one direction within the non-local subspace $\mathrm{NL}_p$. 

The geodesic equation can be easily obtained from the Euler-Arnold equation (\ref{geodesics}). For the local directions, we get
\begin{equation}
\frac{dy^{\alpha}}{ds}=f_{\beta\gamma}^{ \ \ \alpha}y^{\beta}y^{\gamma}+\sum
_{p}\left( 1+\mu_{p}\right) f_{\beta\dot{\alpha}_{p}}^{\ \ \ \alpha}y^{\beta }y^{
\dot{\alpha}_{p}}+\sum_{p}f_{\dot{\alpha}_{p}\beta}^{\ \ \ \ \alpha}y^{\dot {\alpha}
_{p}}y^{\beta}+\sum_{p,q}\left( 1+\mu_{q}\right) f_{ \dot{\alpha}_{p} \dot{\beta}_{q}}^{\ \ \  \alpha }y^{\dot{\alpha}_{p}}y^{\dot{\beta}_{q}}.
\end{equation}
We can take advantage of the antisymmetry of the structure constant to rewrite this as
\begin{equation}
\frac{dy^{\alpha}}{ds}=\sum_{p}\mu_{p}f_{\beta\dot{\alpha}
_{p}}^{\ \ \alpha}y^{\beta}y^{\dot{\alpha}_{p}}+\frac{1}{2}\sum_{p\ne q}\left(\mu_{q}-\mu_p\right)
f_{\dot{\alpha}_{p}\dot{\beta}_{q}}^{\ \ \ \alpha}y^{\dot{\alpha}_{p}}y^{\dot{%
\beta }_{q}}.  \label{EA-local}
\end{equation}
Similarly, for a $p$-hard direction we get 
\begin{equation}
\left( 1+\mu_{p}\right) \frac{dy^{\dot{\alpha}_{p}}}{ds}=\sum_{q}\mu _{q}f_{
\alpha \dot{\beta}_{q}}^{\ \ \dot{\alpha}_{p}}y^{\alpha}y^{\dot{\beta}%
_{q}}+\frac{1}{2}\sum_{q\ne r}\left( \mu_{r}-\mu_q\right) f_{\dot{\beta}_{q}\dot{\gamma}_{r}}^{\ \
\dot{\alpha}_{p}}y^{\dot{\beta}_{q}}y^{\dot{\gamma}_{r}}.
\label{EA-nlocal}
\end{equation}
Here it is clear that the differences of penalties allow hard directions to act as sources for the easy sector.

As done above, we consider now perturbations around the constant solution as
\bea
\label{perturbations}
y^{\alpha}\left(s\right)  = v^{\alpha}+\delta y^{\alpha}\left(s\right), \qquad
y^{\dot{\alpha}_{p}}\left(s\right) = \delta y^{\dot{\alpha}_{p}}\left(s\right).
\eea
The linear equations for the perturbed solutions are the Jacobi equations, which in this case can be written as
\begin{align}
\label{{jacobiequationsHAL}}
\frac{d\delta y^{\alpha}}{ds}  = -i\sum_{p}\mu_{p}\mathrm{C}_{\dot{\alpha}_p}^{\ \alpha}\delta y^{\dot{\alpha}_{p}}, \qquad 
\frac{d\delta y^{\dot{\alpha}_{q}}}{ds} & = -i\sum_{p}\frac{\mu_{p}}{1+\mu_q}\mathrm{C}_{\dot{\alpha}_p}^{\ \dot{\alpha}_q}\delta y^{\dot{\alpha}_{p}}, 
\end{align}
with $p=1,\ldots ,D$. The coefficients on the right-hand sides are the matrix elements of the (super)operator $\mathrm{C}: T[SU(2^N)] \to T[SU(2^N)]$ defined as
\begin{equation}
    \mathrm{C}(X)\equiv  [v,X], \qquad v=v^{\alpha}T_{\alpha},
    \label{Coperator}
\end{equation}
which reads explicitly
\begin{equation}
    \mathrm{C}_{J}^{\ I}= i v^{\alpha}f_{\alpha J}^{\ \ I}\label{matrixelements}.
\end{equation}
As a sanity check, notice that if all cost factors are taken to be equal, we recover the previous case \eqref{jacobieqs}.

\subsubsection{Solution for $D=2$}

To keep things simple we focus here on the case of two degrees of non-locality. We get a set of two coupled equations for the non-local velocities:
\begin{align}
\label{nonlocaleq}
    \frac{d\delta y^{\dot{\alpha}_1}}{ds} &=
    \frac{-i\mu_1}{1+\mu_1}\mathrm{C}^{\dot{\alpha}_1}_{\dot{\beta}_1}
    \delta y^{\dot{\beta}_1}- \frac{i\mu_2}{1+\mu_1}
\mathrm{C}^{\dot{\alpha}_1}_{\dot{\beta}_2} \delta
    y^{\dot{\beta}_2}, \cr
    \frac{d \delta y^{\dot{\alpha}_2}}{ds} &=
    \frac{-i\mu_1}{1+\mu_2}\mathrm{C}^{\dot{\alpha}_2}_{\dot{\beta}_1}
    \delta y^{\dot{\beta}_1}-
    \frac{i\mu_2}{1+\mu_2}\mathrm{C}^{\dot{\alpha}_2}_{\dot{\beta}_2}
    \delta y^{\dot{\beta}_2}.
\end{align}

It is now convenient to introduce a new basis $\{\tilde{T}_{\dot{\alpha}_1},\tilde{T}_{\dot{\alpha}_2} \}$ in the hard directions to diagonalize $\mathrm{C}_{\dot{\beta}_1}^{\ \dot{\alpha}_1}$ and $\mathrm{C}_{\dot{\beta}_2}^{\ \dot{\alpha}_2}$, resulting in\footnote{Here and in the following there is no sum over repeated indices involving the eigenvalues $\lambda^{(\dot \alpha_p)}$.} 
\begin{equation}
    \mathrm{C}_{\dot{\beta}_1}^{\ \dot{\alpha}_1}
    \delta\tilde{y}^{\dot{\beta}_1}=
\lambda^{(\dot{\alpha}_1)}\delta\tilde{y}^{\dot{\alpha}_1}, 
    \qquad   
    \mathrm{C}_{\dot{\beta}_2}^{\ \dot{\alpha}_2}
\delta\tilde{y}^{\dot{\beta}_2}=\lambda^{(\dot{\alpha}_2)}
    \delta\tilde{y}^{\dot{\alpha}_2},
\end{equation}
where the matrix elements in the new basis are given by
\begin{equation}
\label{newC}
\mathrm{C}^{\dot\beta_q}_{\dot\alpha_p}\equiv \frac{1}{2^{N}}
\Tr\left(\tilde{T}_{\dot\beta_q}^{\dagger}[v,\tilde{T}_{\dot\alpha_p}]
    \right).
\end{equation}
We are now in the position to solve the Jacobi equations exactly. The velocities associated with the second level of non-locality can be formally expressed in terms of the first-level velocities as
\begin{equation}
    \delta \tilde y^{\dot{\alpha}_2}(s)= \delta \tilde y^{\dot{\alpha}_2}(0)
    e^{-iM_{\dot{\alpha}_2}s}
     - \frac{i\mu_1}{1+\mu_2} \sum_{\dot{\beta}_1}
    \mathrm{C}^{\ \dot{\alpha}_2}_{\dot{\beta}_1}\int^{s}_0
    ds'\, \delta
    \tilde y^{\dot{\beta}_1}(s')e^{-iM_{\dot{\alpha}_2}
    (s-s')},
\end{equation}
where $M_{\dot\alpha_q} = \lambda^{(\dot{\alpha}_q)} \mu_q / (1 + \mu_q)$. Substituting the expression above in \eqref{nonlocaleq} yields the following integro-differential equation 
\begin{align}
    \frac{d\delta \tilde y^{\dot{\alpha}_1}}{ds}  = 
    -iM_{\dot \alpha_1} \delta
    \tilde y^{\dot{\alpha}_1} & - \frac{i\mu_2}{1+\mu_1}\sum_{\dot{\beta}_2}
    \mathrm{C}^{\ \dot{\alpha}_1}_{\dot{\beta}_2} \delta
    \tilde y^{\dot{\beta}_2}(0)e^{-iM_{\dot{\beta}_2}s} \cr
    & -\frac{\mu_1\mu_2}{(1+\mu_1)(1+\mu_2)}\sum_{\dot{\beta}_1,\dot{\beta}_2}\mathrm{C}^{\ \dot{\alpha}_1}
    _{\dot{\beta}_2}\mathrm{C}^{\ \dot{\beta}_2}_{\dot{\beta}_1}
    \int^{s}_0 ds' \delta \tilde y^{\dot{\beta}_1}(s')
    e^{-iM_{\dot{\beta}_2}
    (s-s')}.
\end{align}
This equation can be handled more easily in Laplace space \cite{volterra1930theory}, since the integral term is a convolution. Thus, let us consider the Laplace transform $\mathcal{L}\left( \delta y(s)\right)=\delta Y(x)$, which satisfies the following algebraic equation
\begin{align}
    \sum_{\dot{\beta}_1}\mathcal{D}^{\ \dot{\alpha}_1}_{\dot{\beta}_1}
    \delta Y^{\dot{\beta}_1}(x) = \delta \tilde y^{\dot{\alpha}_1}(0)
    - \frac{i\mu_2}{1+\mu_1}\sum_{\dot{\beta}_2}\frac{\mathrm{C}_{\dot{\beta}_2}
    ^{\ \dot{\alpha}_1}\delta \tilde y^{\dot{\beta}_2}(0)}
{x+iM_{\dot{\beta}_2}},\label{algebraiceq}
\end{align}
where
\begin{equation}
    \mathcal{D}^{\ \dot{\alpha}_1}_{\dot{\beta}_1}\equiv
    \left(x+iM_{\dot{\alpha}_1}\right)
\delta^{\dot{\alpha}_1}_{\dot{\beta}_1}+
    \Sigma^{\ \dot{\alpha}_1}_{\dot{\beta}_1}, 
    \qquad
    \Sigma^{\ \dot{\alpha}_1}_{\dot{\beta}_1}
    \equiv \frac{\mu_1\mu_2}{(1+\mu_1)(1+\mu_2)}\sum_{\dot{\beta}_2}
    \frac{\mathrm{C}^{\ \dot{\alpha}_1}_{\dot{\beta}_2}
    \mathrm{C}^{\ \dot{\beta}_2}_{\dot{\beta}_1}}
{x+iM_{\dot{\beta}_2}}.\label{defmatrices}
\end{equation}
The equation (\ref{algebraiceq}) is algebraic and it admits a formal solution, provided that the matrix $\mathcal{D}^{\ \dot{\alpha}_1}_{\dot{\beta}_1}$ is invertible:
\begin{equation}
    \delta Y^{\dot{\alpha}_1}(x)= \sum_{\dot \beta_1}
    \left(\mathcal{D}^{\ \dot{\alpha}_1}_{\dot{\beta}_1}\right)^{-1}
    \left( \delta \tilde y^{\dot{\beta}_1}(0)-
    \frac{i\mu_2}{1+\mu_1}\sum_{\dot\beta_2}\frac{\mathrm{C}_{\dot{\beta}_2}
    ^{\ \dot{\beta}_1}\delta \tilde y^{\dot{\beta}_2}(0)}
    {x+iM_{\dot{\beta}_2}}\right).
    \label{laplacedeltay1}
\end{equation}
The term $\Sigma^{\ \dot{\alpha}_1}_{\dot{\beta}_1}$ encodes the influence of the Jacobi fields associated with the most non-local subspace, $\mathrm{NL}_2$, on the subspace $\mathrm{NL}_1$. To obtain the explicit solution, one must invert the Laplace transform. Depending on the form of $\Sigma^{\ \dot{\alpha}_1}_{\dot{\beta}_1}$, this can be a challenging task.

The equation for the local components can be easily integrated, since it depends only on the non-local projections. The final solution is the sum of the components from each subspace
\begin{equation}
    \delta y(s)=\delta y_\textrm{L}(s)+\delta y_{\textrm{NL}_1}(s)+\delta y_{\textrm{NL}_2}(s),
\end{equation}
where
    \begin{align}
    \label{projectionsy}
        \delta y_\textrm{L}(s)&=\frac{1}{2^N}\sum_{\alpha}\mathrm{Tr}\left(\delta y(s)T_{\alpha}\right)T_{\alpha},\cr
        \delta y_{\textrm{NL}_1}(s)&=\frac{1}{2^N}\sum_{\dot{\alpha}_1}\mathrm{Tr}\left(\delta y(s)T_{\dot{\alpha}_1}\right)T_{\dot{\alpha}_1},\cr
        \delta y_{\textrm{NL}_2}(s)&=\frac{1}{2^N}\sum_{\dot{\alpha}_2}\mathrm{Tr}\left(\delta y(s)T_{\dot{\alpha}_2}\right)T_{\dot{\alpha}_2}.
    \end{align}
The traces in the expressions above act as projectors to the respective spaces, as a consequence of the orthogonality condition $\mathrm{Tr}(T_IT_J)=2^N\delta_{IJ}$ satisfied by the generators.


\subsubsection{Location of conjugate points}
\label{locationofCP}

Now we investigate the occurrence and location of conjugate points in some simple situations when two different cost factors are present. 

As a first example, let us consider elements of the non-local subspaces, $w_1 \in \mathrm{NL}_{1}$ and $w_2 \in \mathrm{NL}_{2}$ such that
\begin{equation}
[v,w_1]=\lambda_{1}\,w_1, \qquad [v,w_2]=\lambda_{2}\,w_2 .
\label{assumptions}
\end{equation}
In other words, there exists a $p$–hard direction that diagonalizes the operator $\mathrm{C}(X)=[v,X]$. This implies that the off-diagonal blocks vanish, $\mathrm{C}^{\ \dot\alpha_p}_{\dot\beta_{q}}=0$ for $p\neq q$ and that $\Sigma^{\ \dot\alpha_1}_{\dot\beta_1}=0$.  It then follows that the Jacobi equations decouple in this limit, and their solutions are given by
\begin{equation}
\delta y^{\alpha}(s) = \textrm{const.}, \qquad
\delta y^{\dot{\alpha}_p}\left( s\right) =\delta y^{\dot{\alpha}_{p}}\left( 0\right) \exp\left(-i{\frac{\mu _{p} \lambda
^{\left(p\right) } }{1+\mu _{p}} }s\right).
\end{equation}

Once the Jacobi equations are solved, one can demonstrate that the direction $w_p$ is an eigenvector of the Jacobi operator $\mathbf{Y}_{\mu}$. To see this, take the initial variation $\delta y_{\textrm{NL}_p}(0)=w_p$. Then
\begin{equation}
\mathbf{Y}_{\mu}[w_p]
=\int_0^{1}\! ds\; e^{i v  s}\,w_p\,e^{-i v s}\,
\exp\!\left(-i\frac{\mu_p \lambda^{(p)}}{1+\mu_p} s\right).
\end{equation}
Using the assumption in (\ref{assumptions}), we get a nice simplification for the integral above, which evaluates to
\begin{equation}
\mathbf{Y}_{\mu }\left[ w_{p}\right] =\phi \left( \frac{\lambda ^{\left( p\right) }}{ 1+\mu _{p}}\right) w_{p}. 
\end{equation}
The conjugate points associated with the $p$-hard direction are determined by the zeros of $\phi\left(\frac{\lambda^{(p)}}{1+\mu_p}\right)$. To connect this geometric analysis to evolution in physical time, we use the dictionary in \eqref{dictionary}. Under this identification, the constant geodesic generator is mapped to a time-independent physical Hamiltonian, and the geodesic is mapped to the unitary time evolution $U(t)=e^{-iHt}$, while the geometric spectral quantities are promoted to their time-rescaled counterparts, $\lambda^{(p)} \to t\lambda^{(p)}$. This provides a natural definition of conjugate times in the physical setting:
\begin{equation}
t_{n}^{(p)}=\frac{2\pi n}{\lambda^{(p)}}\,(1+\mu_p),\qquad n\in\mathbb{Z}.
\end{equation}
The conjugate times are shifted by the cost factor $\mu_p$ and are inversely proportional to the eigenvalue $\lambda^{(p)}$. In particular, there are no conjugate points when $\lambda^{(p)}=0$. Moreover, for fixed $\lambda$ and $n$, larger cost factors delay conjugate times: if $\mu_p>\mu_q$, then $t_n^{(p)}>t_n^{(q)}$.

As a second example, we assume mixing between hard directions, that is,
\begin{equation}
\left[ v,w_{q}\right] =\omega_q {}^{p}w_{p}, \qquad  p\neq q=1,2.
\end{equation}%
The coefficients $\omega_q {}^{p}>0$ couple the directions $w_{p}$ and $w_{q}$, which belong to different subspaces. Due to the Hermiticity of the matrices $C_J{}^{I}$, the couplings are symmetric,
$\omega_1 {} ^{2}=\omega_2{}^{1}\equiv \omega.$
In fact, we have $C_{\dot\alpha_i}^{\ \dot\beta_i}=0$ and $\Sigma_{\dot\alpha_1}^{\ \dot\beta_1}= (K^2/x)\delta^{\dot\beta_1}_{\dot\alpha_1}$. Therefore, the Laplace transform in \eqref{laplacedeltay1} becomes
\begin{equation}
\delta Y^{\dot\alpha_1}(x)=\frac{\delta y^{\dot\beta_1}(0)}{x+K^2/x}\delta^{\dot\alpha_1}_{\dot\beta_1}-\frac{i\mu_2}{1+\mu_1}\frac{\delta y^{\dot\beta_2}(0)}{x^2+K^2}\omega_{\dot\beta_2}^{\dot\alpha_1}.
\end{equation}
It can be inverted to find the velocities. It is worth noting that, in the above equation, the indices $\dot\alpha_1,\ \dot\beta_2$ should be understood as fixed indices in the non-local subspace. Thus, the Jacobi equations decouple into independent pairs. With appropriate initial conditions and applying the  dictionary $(v,\omega)\mapsto (Ht,t\omega)$ in \eqref{dictionarydefs}, the solution can be written as
\begin{eqnarray}
\delta y^{\dot{\alpha}_{1}} &=&\delta y^{\dot{\alpha}_{1}}\left( 0\right)
\cos \left( tKs\right) -\frac{i}{\epsilon }\, \delta y^{\dot{\alpha}%
_{2}}\left( 0\right) \sin \left( tKs\right) , \cr
\delta y^{\dot{\alpha}_{2}} &=&\delta y^{\dot{\alpha}_{2}}\left( 0\right)
\cos \left( tKs\right) -i\epsilon\,  \delta y^{\dot{\alpha}_{1}}\left( 0\right)
\sin \left( tKs\right) ,
\end{eqnarray}%
with
\begin{equation}
\epsilon =\sqrt{\frac{\mu _{1}\left( 1+\mu _{1}\right) }{\mu _{2}\left(
1+\mu _{2}\right) }},\qquad K=\omega \sqrt{\frac{\mu _{1}\mu
_{2}}{\left( 1+\mu _{2}\right) \left( 1+\mu _{1}\right) }}.
\end{equation}
With these definitions, it is possible to evaluate the integral form of the Jacobi operator. In fact, we obtain a mixing of the two non-local subspaces. Schematically, we have
\begin{equation}
\mathbf{Y}_{\mu }\left[ w_{p}\right] =\mathbf{Y}_{\text{NL}_1}\left[ w_{p}\right]
w_{p}+\mathbf{Y}_{\text{NL}_2}[w_{p}]w_{q},\qquad 
p\neq q=1,2,
\end{equation}
where $\mathbf{Y}_{\text{NL}_q}\left[ w_{p}\right]$ denotes the components of  $\mathbf{Y}_{\mu}\left[ w_{p}\right]$ along the $q$–hard subspace. This means that the subspace spanned by $\{w_{1},w_{2}\}$ is invariant. Therefore, the spectrum can be identified with the eigenvalues of the matrix
\begin{equation}
(\mathbf{Y}_{\mu})_{pq}=\mathbf{Y}_{\mathrm{NL}_p}[w_q].
\label{Yijmatrix}
\end{equation}
To obtain the conjugate times we impose $\det \mathbf{Y}_{\mu}=0$, resulting in
\begin{equation}
\frac{(1-\epsilon)(K-\omega)}{(1+\epsilon)(K+\omega)}
\left|\sin\left(\frac{(K+\omega)t}{2}\right)\right|=
\left|\sin\left(\frac{(K-\omega)t}{2}\right)\right|.
\label{transEqT}
\end{equation}
This equation has a discrete set of solutions $\{t_n\}$. Notice that the presence of two degrees of non-locality introduces two frequencies, $K\pm\omega$, producing a richer structure of conjugate times. We parametrize the cost factors in polar coordinates:
\begin{equation}
\mu_1=\mu\cos\theta,\qquad \mu_2=\mu\sin\theta,\label{polarparametrization}
\end{equation}
where $\mu\in\mathbb{R}^+$ measures the overall cost scale, while $\theta\in[\pi/4,\pi/2]$ controls the anisotropy between the two non-local sectors, with $\mu_2>\mu_1$. In these coordinates, the symmetric case is obtained at $\theta=\pi/4$, for which $\epsilon=1$. In this limit, \eqref{transEqT} reduces to the well-known single cost factor behavior:
\begin{equation}
t_n=\frac{2\pi n}{K-\omega}\quad \Rightarrow \quad
t_n=\left(1+\mu/\sqrt{2}\right)\frac{2\pi n}{\omega}.
\end{equation}

\begin{figure}[t]
\begin{minipage}{.5\textwidth}
        \centering
        \includegraphics[width=\textwidth]{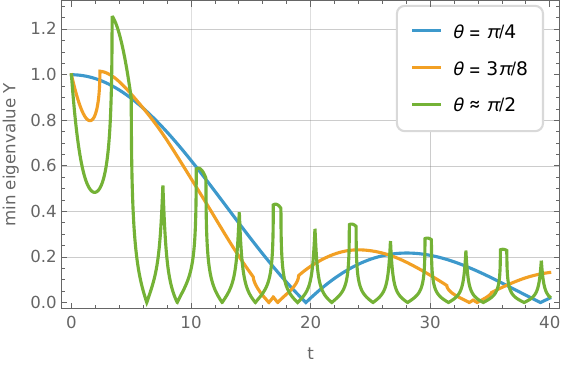}
    \end{minipage}
\hfill 
\begin{minipage}[H]{.5\textwidth}
        \centering
    \includegraphics[width=\textwidth]{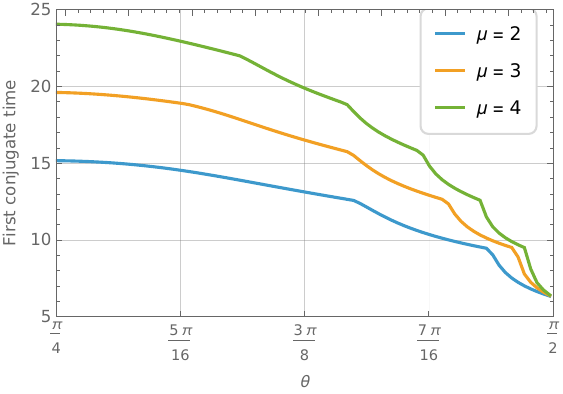}
    \end{minipage}
\caption{
Left: Time evolution of the minimum
 eigenvalue of the Jacobi operator for three anisotropy angles at fixed $\mu=3$ and $\omega=1$. Conjugate times are identified by zero crossings. As $\theta$ increases, the oscillatory frequency increases and zero crossings appear earlier.
Right: First conjugate time $t_{\min}$ as a function of $\theta$ with $\omega=1$. Larger values of $\mu$ delay conjugate times in the symmetric regime, while near $\theta\to\pi/2$ the curves approach $t_{\mathrm{min}} \approx 6.4$.
}
 \label{conjugT}  
\end{figure}

In the general case, the conjugate-time structure is depicted in figure \ref{conjugT}, where we show the evolution of the minimum eigenvalue of the Jacobi operator. Each zero corresponds to one conjugate time. The effect of multiple cost factors is clear: increasing the anisotropy angle leads to earlier conjugate times, whereas increasing $\mu$ shifts conjugate points to later times. In the limit $\theta\to\pi/2$, when $\mathrm{NL}_1$ effectively approaches the local sector $\mathrm{L}$, the first conjugate time becomes approximately independent of $\mu$.


\section{Complexity of a qubit with two cost factors}
\label{sec:1qubit}
We now apply the generic results obtained above to the simplest possible system: a single qubit. The associated group manifold is $SU(2)$, it is generated by the three Pauli matrices $T_{i}=\sigma_{i}$ ($i=1,2,3$) and can be split into one easy direction, $\sigma_{1}$, and two hard directions, $\sigma_{2}$ and $\sigma_{3}$ with cost factors $\mu_{2}$ and $\mu_{3}(\neq\mu_{2})$, respectively.

The geodesic equations \eqref{EA-local} in this case are given by
\bea
\frac{dy^1}{ds} &=&  2\left(\mu_3 - \mu_2\right)y^2 y^3, \cr
\frac{dy^2}{ds} &=&  -\frac{2\mu_3}{1 + \mu_2}y^3 y^1, \cr
\frac{dy^3}{ds} &=&  \frac{2\mu_2}{1 + \mu_3}y^1 y^2. \label{oneqbitEA}
\eea
Notice that if $\mu_{2}=\mu_{3},$ the first equation becomes trivial, resulting in a linear system for the velocities in the hard direction, $y^{2}$ and $y^{3}$. This is a qubit with just one cost factor, which is equivalent, up to redefinitions, to the case considered in \cite{Brown_2019, Balasubramanian_2020}.

On the other hand, when $\mu_2 \neq \mu_3$, the Euler-Arnold equations for the single-qubit case form a genuinely non-linear, but still integrable, system. Notice that the derivative of each quadratic term $(y^i)^2$ is proportional to $y^1 y^2 y^3$. This suggests looking for a quadratic quantity of the form
\begin{equation}	I=a_1\left(y^1\right)^2+a_2\left(y^2\right)^2+a_3\left(y^3\right)^2,\qquad a_i\in\mathbb{R}.
\end{equation}
Requiring this quantity to be $s$-independent and using the Euler-Arnold equations, we obtain the following algebraic 
condition on the coefficients $a_i$
\begin{equation}
\alpha a_1+\beta a_2+\gamma a_3=0, \label{algebriceq}
\end{equation}
with
\begin{equation}
	\alpha  =2\left(  \mu_{3}-\mu_{2}\right)  ,\qquad
	\beta  =-\frac{2\mu_{3}}{1+\mu_{2}},\qquad
	\gamma  =\frac{2\mu_{2}}{1+\mu_{3}}.
\end{equation}
Since \eqref{algebriceq} is a single linear constraint on three coefficients, its solution space is two-dimensional. Hence, there are two linearly independent conserved quadratic quantities. For instance, setting $a_3=0$ and $a_1=1$, we obtain
\begin{equation}
a_2=-\frac{\alpha}{\beta} = \frac{\left(\mu_3-\mu_2\right)\left(1+\mu_2\right)}{\mu_3},
\end{equation}
which yields the first conserved quantity
\begin{equation}
I_1=
\left(y^1\right)^2
-\frac{\alpha}{\beta}
\left(y^2\right)^2.
\end{equation}
Similarly, setting $a_2=0$ and $a_1=1$ gives $a_3=-\alpha/\gamma$,
leading to the second conserved quantity
\begin{equation}
I_2=
\left(y^1\right)^2
-\frac{\alpha}{\gamma}
\left(y^3\right)^2.
\end{equation}

Geometrically, these conserved quantities define two surfaces in the three-dimensional space $(y^1,y^2,y^3)$, and the dynamics is confined to their intersection. Assuming $\mu_3>\mu_2>0$, we have $\alpha>0$ and $\beta<0$, and therefore the first conserved quantity satisfies $I_1\geq 0$ and defines an elliptic cylinder. On the other hand, the second conserved quantity $I_2$ defines a hyperbolic cylinder, with the case $I_2=0$ corresponding to its degenerate limiting surface.

\begin{figure}[t]
    \centering
    \includegraphics[width=0.72\textwidth]{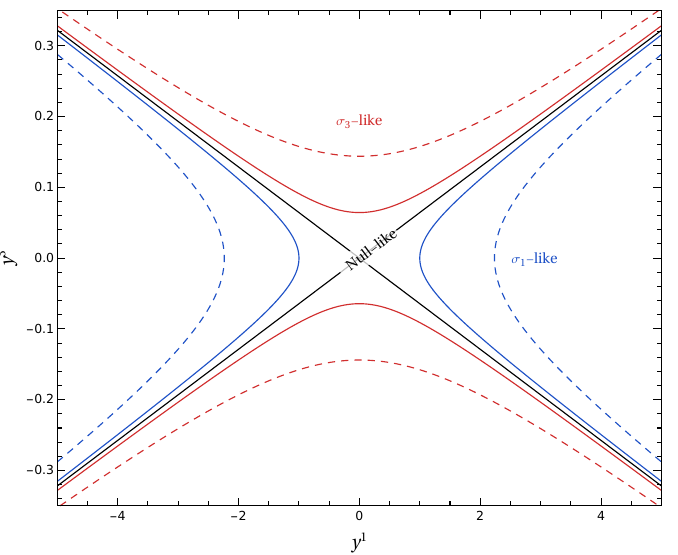}
    \caption{Geometric classification of the level sets of the conserved quantity $I_2$ in the $(y^1,y^3)$ plane. The blue curves correspond to $I_2>0$ and define the $\sigma_1$-like sector, whereas the red curves correspond to $I_2<0$ and define the $\sigma_3$-like sector. Solid and dashed curves represent different values of $|I_2|$, namely $I_2=\pm 1$ and $I_2=\pm 5$, respectively. The black lines correspond to $I_2=0$ and separate the two sectors. We use the polar parametrization $\mu_2=\mu\cos(\theta)$ and $\mu_3=\mu\sin(\theta)$, with $\mu=50$ and $\theta=1.4$.}
    \label{I2regimes}
\end{figure}

Using these conserved quantities, we can reduce the system \eqref{oneqbitEA} to a single equation for the component $y^1$, which reads
\begin{equation}
\left( \frac{d y^{1}}{ds}\right)  ^{2}=\lambda^{2}\left(  I_{1}-\left(y^{1}\right)  ^{2}\right)  \left(  \left(  y^{1}\right)  ^{2}-I_{2}\right),
\qquad\lambda=2\sqrt{\frac{\mu_{2}\mu_{3}}{\left(  1+\mu_{2}\right) \left(1+\mu_{3}\right)  }}.
\label{reducedeq}
\end{equation}
We can plot $I_2$ in the $(y^1,y^3)$ plane, see figure \ref{I2regimes}. 
There are three different cases, corresponding to the possible different signs of $I_2$, which we denote as follows: 
\begin{align*}
	& I_2 >0: \quad \text{$\sigma_1$-like},\\
	& I_2 =0: \quad \text{null-like},\\
	& I_2 <0: \quad \text{$\sigma_3$-like}.
\end{align*}

Depending on the sign of $I_2$, the integral associated with \eqref{reducedeq} can be expressed in terms of different Jacobi elliptic functions. A convenient way to organize the solutions for both regimes, $I_2\gtrless0$, is to introduce the projectors $p_\pm = (1\pm\mathrm{sign}(I_2))/2$. The solution can then be written in compact form
\begin{equation}
	y_1(s)=\mathcal{A}\left[\pm p_+\, \mathrm{dn}(u,\varepsilon)+p_-\, \varepsilon\,\mathrm{cn}(u,\varepsilon)\right], \qquad u= u_0+\lambda\mathcal{A}s,
    \label{soly1}
\end{equation}
where the choice $\pm$ in the first term reflects the fact that for $\sigma_1$-like motion we have two possible disconnected hyperbolic trajectories in the $\sigma_1$ direction, see figure \ref{I2regimes}. Moreover $u_0 \in \mathbb{R}$ is the initial elliptic phase, and the amplitude $\mathcal{A}$ and the eccentricity $\varepsilon$ of the motion are given by
\begin{equation}
	\mathcal{A}=\sqrt{I_1-p_-\, I_2},\qquad \mathrm{\varepsilon}=(1-I_2/I_1)^{\mathrm{sign}(I_2)/2}.
\end{equation}
We stress that $I_1$ and $I_2$ remain constant throughout the motion. Therefore, once $y^1(s)$ is known from \eqref{soly1}, the remaining components $y^2(s)$ and $y^3(s)$ can be reconstructed from the conserved quantities.

Finally, we can express the complexity in \eqref{relativecomplexity} directly in terms of the conserved quantities.
For a metric with two cost factors, the complexity reads
\begin{equation}
    \mathcal{C}[U_{\mathrm{target}}]=\min\int^{1}_0ds \sqrt{(y^1)^2+(1+\mu_2)(y^2)^2+(1+\mu_3)(y^3)^2}.
\end{equation}
The quadratic form inside the square root can be rewritten in terms of the constant of motion,
leading to the following convenient expression for the complexity of an arbitrary single-qubit target gate
\begin{equation}
\mathcal{C}[U_{\mathrm{target}}]
=
\min_{u_0,I_1,I_2}
\sqrt{
\frac{\mu_3 I_1-\mu_2 I_2}
{\mu_3-\mu_2}
}.
\label{complexityinvariants}
\end{equation}
Here, minimization is performed on the set of admissible integration constants for which the corresponding geodesic satisfies $U(1)=U_{\mathrm{target}}$. The reconstruction of the path on $SU(2)$ is discussed in the next subsection.

\subsection{Numerical scheme for the Dyson equation}

As explained in section \ref{sec:theory}, the geodesic connecting the identity to the target operator must satisfy the Dyson equation
\begin{equation}
\frac{dU(s)}{ds}=-iy^i(s)T_i\,U(s),
\label{dysoneq}
\end{equation}
together with the boundary conditions
\begin{equation}
U(0)=\mathbbm{1},
\qquad
U(1)=U_{\rm target}(t)=e^{-iHt}.
\label{eq:BC}
\end{equation}

The generator along the geodesic is determined by the Euler-Arnold solution $y^i(s)$. Even though the functions $y^i(s)$ are known analytically from the previous section, the corresponding Dyson equation can be complicated to solve analytically. We therefore integrate \eqref{dysoneq} numerically using the exact Euler-Arnold solution as input. 

It is sufficient to consider only the first column in the matrix $U(s)$. Indeed, a generic element of $SU(2)$ can be written in terms of only two complex components, since $U_{22}=U_{11}^{*}$ and $U_{12}=-U_{21}^{*}$. Therefore, the Dyson equation \eqref{dysoneq} reduces to the coupled system
\begin{align}
\frac{dU_{11}}{ds}
&=
-i\,y^3 U_{11}
-i\left(y^1-i y^2\right)U_{21},
\cr
\frac{dU_{21}}{ds}
&=
-i\left(y^1+i y^2\right)U_{11}
+i\,y^3 U_{21},
\end{align}
with the initial conditions
\begin{equation}
U_{11}(0)=1,
\qquad
U_{21}(0)=0.
\end{equation}
This initial-value problem is numerically integrated over the interval $s\in[0,1]$ using the \texttt{solve\_ivp} routine from \texttt{SciPy}. For a given target and time $t$, the remaining boundary condition at $s=1$ is imposed through a numerical minimization procedure. In particular, the integration constants obtained from the Euler-Arnold solution, which may be taken as $(I_1,I_2,u_0)$, together with the corresponding discrete branch labels, are varied so as to satisfy
\begin{equation}
U_{11}(1)=U_{11}^{\rm target}(t),
\qquad
U_{21}(1)=U_{21}^{\rm target}(t).
\label{eq:shootingcondition}
\end{equation}
Numerically, this is implemented by minimizing the residual
\footnote{The minimization is performed using the \texttt{least\_squares} from \texttt{SciPy}.}
\begin{equation}
\mathrm{RES}
=
\sqrt{
\left|U_{11}(1)-U_{11}^{\rm target}(t)\right|^2
+
\left|U_{21}(1)-U_{21}^{\rm target}(t)\right|^2
}.
\label{eq:residual}
\end{equation}

Since the boundary-value problem may admit multiple geodesics connecting the identity to the same target operator, the minimization of \eqref{eq:residual} is performed using multiple initial guesses for the integration constants. The procedure is carried out independently in the three dynamical sectors corresponding to the different signs of $I_2$. For each target time $t$, the complexity is then taken to be the minimum among all admissible geodesics found.


\subsection{Time evolution of complexity}

So far, we have developed a numerical strategy to determine geodesics in $SU(2)$ by combining the Euler-Arnold solutions with the boundary conditions in \eqref{eq:BC}. We now consider three different one-qubit Hamiltonians in order to investigate how the time evolution of complexity depends on the multiple cost factors. We compare three qualitatively different situations: evolution purely along the easy direction, evolution generated by a mixture of an easy and a penalized direction, and evolution entirely along the hardest direction.

\begin{figure}[t]
    \centering
    \includegraphics[width=0.72\textwidth]{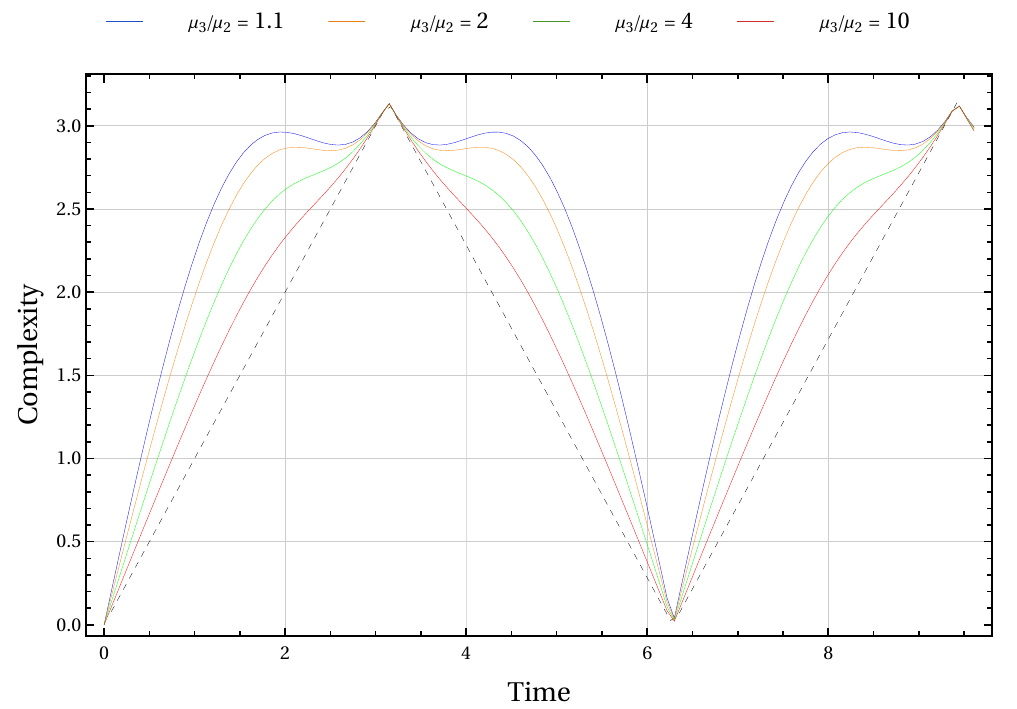}
\caption{Time evolution of the complexity for the target in \eqref{eq:HB} with $J_1=1$, $J_2=0.2$, for different ratios $\mu_3/\mu_2$ with $\mu \approx 200$. The dashed line represents the complexity \eqref{eq:CA}, corresponding to a time evolution generated only by the easy direction ($J_2=0$ case).}
\label{fig:new2}
\end{figure}

We first consider the case of an Hamiltonian along the easy direction:\footnote{We thank the anonymous referee of PRD for spotting an inconsistency in the previous treatment of this case and for prompting us to revisit the computations of this section.}
\begin{equation}
    H_A = J_1 \sigma_1 ,
    \qquad
    U_A(t)=e^{-iJ_1 \sigma_1 t}.
\end{equation}
In this case, the evolution of complexity shown by the dashed black line in figure \ref{fig:new2} can be understood easily. Notice that the $s$-independent solution
\begin{equation}
    y^1(s)=J_1 t,
    \qquad
    y^2(s)=y^3(s)=0,
\end{equation}
satisfies the Euler-Arnold equations. In fact, this simple solution is contained in the $I_2>0$ family, which corresponds to the degenerate elliptic limit with $\varepsilon=0$. Since the geodesic lies entirely along the easy direction, its length does not depend on either penalty factor. Taking into account the compactness of $SU(2)$, the complexity \eqref{complexityinvariants} can be rewritten as
\begin{equation}
    \mathcal{C}_A(t)=\min_{n\in\mathbb{Z}} \left|J_1 t+2\pi n\right|.\label{eq:CA}
\end{equation}
Hence, $\mathcal{C}_A(t)$ displays the usual sawtooth behavior that grows linearly until the target reaches the antipodal point, after which the geodesic with the opposite orientation becomes shorter~\cite{balasubramanian2021complexity}.

Next, we introduce a contribution from the first penalized direction
\begin{equation}
    H_B = J_1\sigma_1 + J_2\sigma_2,
    \qquad
    U_B(t)=
    e^{-it(J_1\sigma_1+J_2\sigma_2)}.\label{eq:HB}
\end{equation}
In contrast to the previous case, the constant solution
$y^{1,2}(s)=J_{1,2}t$, with $y^3=0$, is not an exact solution of the Euler-Arnold equations. Therefore, in this case, the geodesics can explore the hardest direction. The numerical results are displayed in figure \ref{fig:new2} and show linear growth at early times
\begin{equation}
    \mathcal C_B(t)
    \simeq t \sqrt{J_1^2+(1+\mu_2)J_2^2}.
\end{equation}
After that, the complexity presents a pronounced deformation for smaller values of the ratio $\mu_3/\mu_2$. In particular, we observe a change in monotonicity, with the appearance of a local maximum for $\mu_3/\mu_2 \lesssim 2$. The minimizing geodesics remain in the $I_2>0$ sector throughout the evolution, corresponding to the $\sigma_1$-like family. However, at intermediate times, the ratio $I_2/I_1$ becomes very small, indicating that the geodesics approach the null-like family $I_2=0$ without crossing it. Moreover, close to the antipodal point, all curves converge to the value associated with the target solely along the easy direction $H_A$. This can be easily understood, since this point can be reached using only the easy direction $\sigma_1$. Therefore, the influence of multiple cost factors is more evident at intermediate times.

\begin{figure}[t]
    \centering
    \includegraphics[width=0.72\textwidth]{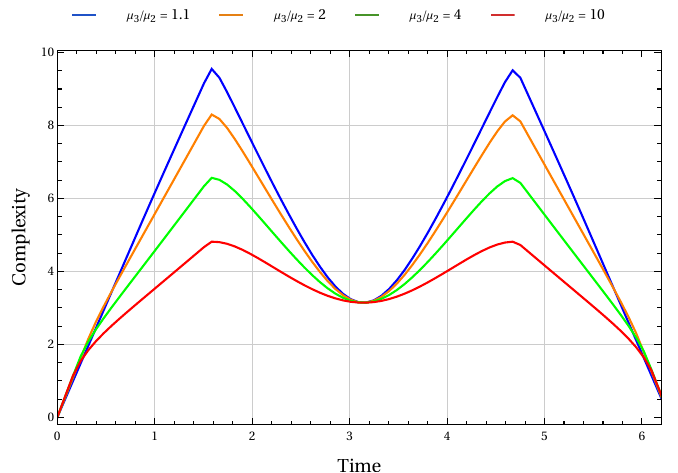}
    \caption{Evolution of the complexity for the target \eqref{eq:HC}, for different ratios $\mu_3/\mu_2$, with $J_3=1$ and $\mu\simeq 50$.}
    \label{fig:new3}
\end{figure}

Finally, we consider a Hamiltonian entirely aligned along the hardest direction
\begin{equation}
    H_C = J_3\sigma_3,
    \qquad
    U_C(t)=e^{-iJ_3t\sigma_3}.
    \label{eq:HC}
\end{equation}
In this case, the constant solution $y^3(s)=J_3t$, with $y^1=y^2=0$, is an exact solution of the Euler-Arnold equations and belongs to the $\sigma_3$-like family. Using \eqref{complexityinvariants}, the corresponding complexity is
\begin{equation}
    \mathcal C_C(t)
    =
    \sqrt{1+\mu_3}\,J_3 t.
\end{equation}
However, this linear growth fails to be the global minimum at early times. The numerical results shown in figure \ref{fig:new3} present a qualitatively different behavior from the previous cases, with a crossing between the different geodesic sectors given by the signs of $I_2$. After the initial linear growth along the constant $\sigma_3$ geodesic, the easy directions start to compete strongly in determining the minimizing geodesic, producing a complexity profile with a pronounced dependence on the cost factors. As a consequence, the complexity develops a pronounced maximum at intermediate times and then decreases as the target approaches the antipodal point, where the minimizing geodesic becomes $\sigma_1$-like, similarly to the previous case.


\section{Conjugate points in the SYK model}
\label{sec:SYK}

So far we have considered a one-qubit system. Now we move on to systems with more qubits and we do so in the context of the SYK model \cite{Sachdev_1993, Sachdev_2024, orman2024quantumchaossparsesyk}. This model describes $N$ Majorana fermions $\psi_i$ ($i=1,2,\ldots,N$) with Hamiltonian
\begin{equation}
    H= c_q\sum_{i_1>i_2>\ldots >i_q}J_{i_1i_2\cdots i_q}\psi_{i_1}\psi_{i_2}\ldots\psi_{i_q}
    \label{SYKH},
\end{equation}
where $q<N$ and the random couplings $J_{i_1i_2\ldots i_q}\in \mathbbm{R}$ are totally antisymmetric and the overall phase $c_q$ is chosen in each case to ensure hermiticity of the Hamiltonian. If we introduce the collective index ${\cal I}=i_1i_2\ldots i_q$, these couplings are drawn from a Gaussian ensemble with zero mean and variance
\begin{equation}
    \langle J_{\cal I}J_{{\cal I}'}\rangle=\delta_{{\cal II}'}\frac{J^2(q-1)!}{N^{q-1}},
\end{equation}
for some constant $J$. 


\subsection{Free SYK model}
\label{syk2}

We start by analyzing the emergence of conjugate points in the quadratic ($q=2$) SYK model, which we call {\it free} in the following:
\begin{equation}
    H_\textrm{free} =  i  \sum_{i > j} J_{ij} \psi_i \psi_j.
    \label{SYKH22}
\end{equation}
We consider the case in which two different cost factors are included. We first compute the explicit solution for the Jacobi fields and then we numerically investigate the spectrum of the Jacobi operator to determine the location of the conjugate points. Having more than one cost factor introduces an internal structure within the non-local subspace, affecting the form of the Jacobi fields, as various non-local components interact through a set of coupled differential equations.

In section \ref{sec:theory}, we discussed the general solution using Laplace transforms for the first level of non-locality and an integral equation for the most non-local components of the Jacobi fields. Following \cite{balasubramanian2021complexity}, we compute here the Jacobi fields explicitly, using the Hamiltonian above. We start with the matrix elements
\begin{equation}
    \mathrm{C}_I^J = -\frac{ i  t}{2^{N/2}} \mathrm{Tr}\left(T^\dagger_J [H_\textrm{free}, T_I]\right), \label{CIJ}
\end{equation}
where the indices $I, J$ run over both local (with indices $\alpha,\beta,\ldots$) and non-local subspaces $\mathrm{NL}_1$ (with indices $\dot\alpha_1,\dot\beta_1,\ldots$) and $\mathrm{NL}_2$ (with indices $\dot\alpha_2,\dot\beta_2,\ldots$). In the $\{T_\alpha,\tilde{T}_{\dot\alpha_{p=1,2}}\}$ basis, the matrix (\ref{CIJ}) becomes block-diagonal
\begin{equation}
    \mathrm{C}_I^J =
    -it\begin{pmatrix}
        \mathrm{C}^{\beta}_{\alpha} & 0 & 0 \\
        0 & \lambda^{(\dot\alpha_1)} \delta^{\dot\beta_1}_{\dot\alpha_1} & 0 \\
        0 & 0 & \lambda^{(\dot\alpha_2)} \delta^{\dot\beta_2}_{\dot\alpha_2}
    \end{pmatrix}.
\end{equation}
The Jacobi equations decouple and the solutions for the two non-local subspaces take the same functional form
\begin{equation}
    \delta y^{\dot{\alpha}_p}(s) = \delta y^{\dot{\alpha}_p}(0)\, 
    e^{-\frac{it\mu_p\lambda^{(\dot\alpha_p)}s}{1+\mu_p}},\qquad p=1,2,
\end{equation}
while the local components remain constant
\begin{equation}
    \delta y^{\alpha}(s) = \delta y^{\alpha}(0).
\end{equation}
As a result, the explicit form of the Jacobi operator $ \mathbf{Y} _{\mu}$ is particularly simple and given by
\begin{align}
    \mathbf{Y}_{\mu}\left[ \delta y(0) \right] 
    = \int_0^1 ds \; e^{ i H_\textrm{free}ts}
    \left( \delta y_\textrm{L}(0) 
    + \sum_{p=1,2} \sum_{\dot{\alpha}_p} \delta y^{\dot{\alpha}_p}(0)
    e^{-\frac{it\mu_p\lambda^{(\dot\alpha_p)}s}{1+\mu_p}} \tilde{T}_{\dot\alpha_p}
    \right) e^{- i H_\textrm{free}ts}.\label{Ysyk}
\end{align}
As discussed in section \ref{locationofCP}, the location of the conjugate points is given by the zero-modes of $ \mathbf{Y} _{\mu}$, which we determine next. 

In the same generator basis, the Jacobi operator $\mathbf{Y}_\mu$ takes the block-diagonal form
\begin{equation}
(\mathbf{Y}_\mu)_{IJ}
=
\mathrm{Tr}\left(
T_I^\dagger\, \mathbf{Y}_\mu[T_J]
\right)= \left(
\begin{array}{c c c}
Y^{\mathrm{L}}_{\alpha\beta} & 0 & 0 \\ 
0 & Y^{\mathrm{NL}_1}_{\dot\alpha_1\dot\beta_1} & 0 \\
0 & 0 & Y^{\mathrm{NL}_2}_{\dot\alpha_2\dot\beta_2}
\end{array}
\right).
\label{Yij}
\end{equation}
The non-local blocks are also diagonal, that is,
\begin{equation} 
Y_{\dot\alpha_1\dot\beta_1}^{\textrm{NL}_1} =
\phi\left(\frac{t \lambda^{(\dot{\beta}_{1})}}
{1 + \mu_1}\right) \delta_{\dot\alpha_1\dot\beta_1}, \qquad Y_{\dot\alpha_2\dot\beta_2}^{\textrm{NL}_2} =
\phi\left(\frac{t \lambda^{(\dot{\beta}_{2})}}
{1 + \mu_2}\right) \delta_{\dot\alpha_2\dot\beta_2} ,
\end{equation}
while the local contribution is non-diagonal. However, it can be computed by expanding the generators $T_\alpha$ in the eigenbasis $\{\ket{m}\}$, obtaining
\begin{align}
 Y^{\mathrm{L}} _{\alpha \beta} &= \sum_{m,n,p,q} \mathrm{Tr} \int_{0}^{1} ds \;
c_{qp}^{(\alpha)} c_{mn}^{(\beta)} \left| q \right\rangle \left\langle p \right| 
e^{ i  H_\textrm{free} t s} \left| m \right\rangle \left\langle n \right| 
e^{- i  H_\textrm{free} t s} \cr
&=\sum_{m,n} c_{nm}^{(\alpha)} c_{mn}^{(\beta)}
\phi \left( \Delta_{mn} t \right),
\label{local-partY}
\end{align}
with the coefficients $c^{(\alpha)}_{mn} = \langle m | T_\alpha | n \rangle$ and $\Delta_{mn}=E_m-E_n$, with $H_\textrm{free}| n \rangle=E_n | n \rangle$.

Due to the partially diagonal structure of (\ref{Yij}), we can compute the entire spectrum, which can be organized into an array of $4^{N/2} - 1$ entries in the following schematic form:
\begin{equation}
\Lambda(t) = \left\{
\lambda_{\mathrm{local}}(t),
\phi\left( \frac{t \lambda^{(\dot{\alpha}_1)}}{1 + \mu_1} \right),
\phi\left( \frac{t \lambda^{(\dot{\alpha}_2)}}{1 + \mu_2} \right)
\right\}.
\label{eigenvalues}
\end{equation}
\begin{figure}[t]
\centering
\includegraphics[scale=0.55]{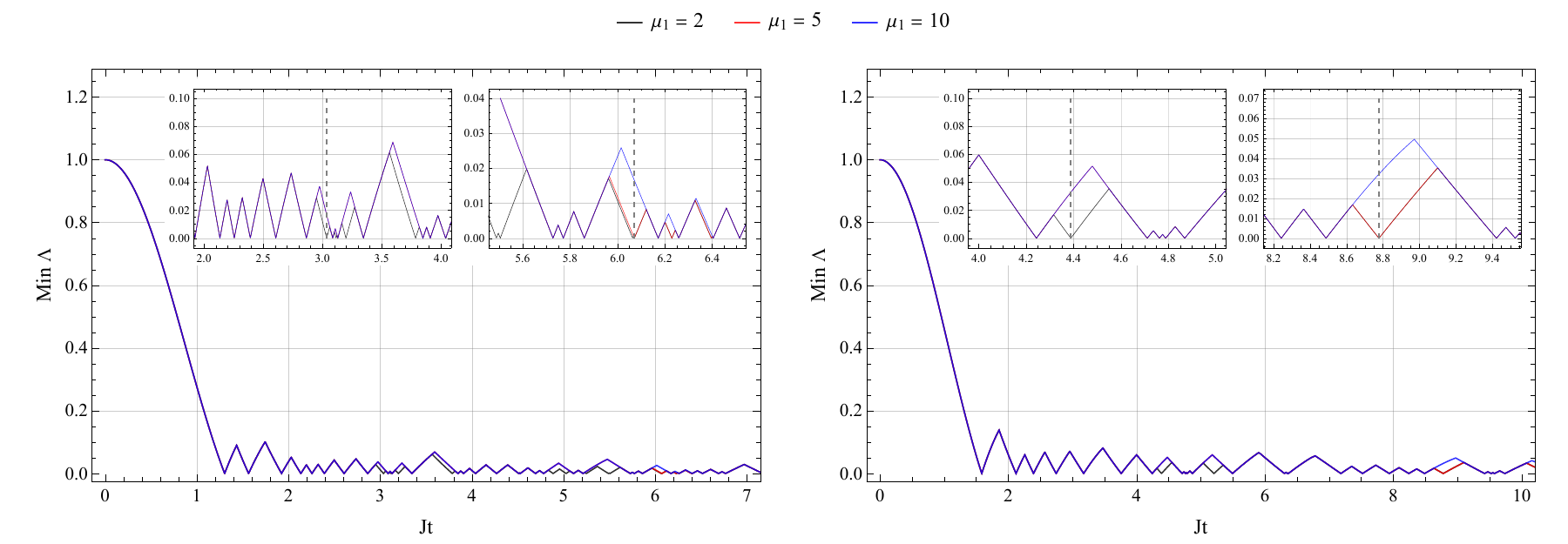}
\caption{Time evolution of the minimum eigenvalue of the Jacobi operator $\mathbf{Y}_\mu$ for different values of $\mu_1$ at fixed $\mu_2=15$. The left panel corresponds to $N=8$, while the right panel corresponds to $N=6$. The insets mark the first visible non-local conjugate times associated with the $\mathrm{NL}_1$ directions.}\label{minYmu1}
\end{figure}
A zero mode occurs at $t_*$ whenever $\Lambda_a(t_*) = 0$ for some component $a$ among the $4^{N/2} - 1$ entries of (\ref{eigenvalues}). In our numerical calculation, we set $N = 6$ and $8$, which correspond to three and four qubits, respectively. We choose the operators $\psi_i$ and $\psi_i \psi_j$ as the local/easy ones. When $N = 6$, the most non-local directions (with cost factor $\mu_2$) consist of operators with four, five, and six Majorana fields, while the less non-local ones (with cost factor $\mu_1$) are built from three fields. For $N = 8$, we set the three- and four-body fermionic operators as the least non-local directions, and the others as the most non-local directions. One can see from (\ref{eigenvalues}) that the non-local conjugate points occur at times
\begin{equation}
t^{\mathrm{NL}_p}_n = 2\pi n (1 + \mu_p) (\lambda^{\dot\alpha_p})^{-1}, \label{nlc}
\end{equation}
where $n$ is an integer that counts the order of the conjugate  points. Since $\mu_2 > \mu_1$, the $n$-th conjugate point  associated with the least non-local directions appears parametrically earlier -- if we suppose that all eigenvalues have the same magnitude -- than the one associated with the most non-local directions, as shown in the figures \ref{minYmu1} and \ref{N8minYmu2}.
\begin{figure}[t]
\centering
\includegraphics[scale=0.8]{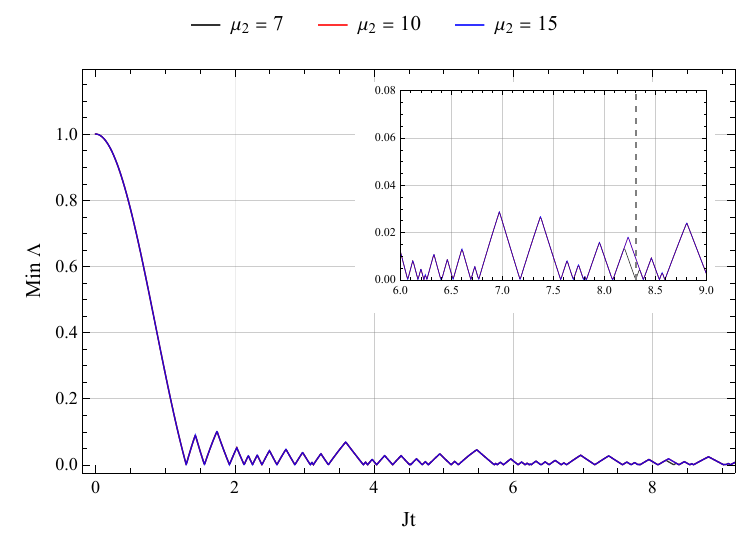}
\caption{Time evolution of the minimum eigenvalue of the Jacobi operator $\mathbf{Y}_\mu$ for $N=8$ and different values of $\mu_2$ at fixed $\mu_1=5$. The inset shows the first visible non-local conjugate time associated with the $\mathrm{NL}_2$ directions, located at $Jt^{\mathrm{NL}_2} \approx 8.3$.}\label{N8minYmu2}
\end{figure}
The first conjugate point associated with the subspace $\mathrm{NL}_i$ is determined by the largest value of $\lambda^{(\dot\alpha_i)}$. For the realization shown in figure \ref{minYmu1}, we have $1/\max\{\lambda^{(\dot\alpha_1)}\} \approx 0.16$ for $N=8$. This yields the first non-local conjugate times $Jt^{\mathrm{NL}_1} \approx 3$, $6.1$, and $11.1$ for different values of $\mu_1$. The first two are visible in figure \ref{minYmu1} (left), while the third is outside the plotted time window. We observe the same pattern for the harder directions shown in figure \ref{N8minYmu2}: since $\max\{ \lambda^{(\dot\alpha_2)} \}\approx \max \{\lambda^{(\dot\alpha_1)}\}$, the first conjugate point associated with $\mathrm{NL}_2$ for $\mu_2=7$ occurs at $Jt^{\mathrm{NL}_2} \approx 8.3$. These results show that penalty factors shift non-local conjugate points to later times, in quantitative agreement with \eqref{nlc}. By contrast, the first zeros in both figures \ref{minYmu1} and \ref{N8minYmu2} do not depend on $\mu_1$ or $\mu_2$, indicating that they originate from the local sector. 


\subsection{Chaotic SYK}

After analyzing the occurrence of conjugate points in the quadratic SYK model, let us now consider SYK Hamiltonians with three- and four-body interaction terms. We call them {\it chaotic}. With the appropriate $c_q$ phases, these Hamiltonians read 
\begin{equation}
H_{\mathrm{SYK3}}=i\sum_{i>j>k}J_{ijk}\psi _{i}\psi _{j}\psi _{k}, \qquad H_{\mathrm{SYK4}}=\sum_{i>j>k>l}J_{ijkl}\psi _{i}\psi _{j}\psi _{k}\psi _{l}.
\label{Hsykchaotic}
\end{equation}
We set $N=8$, which corresponds to 4 qubits.\footnote{The case $N=6$ (3 qubits) does not accommodate an approximately equal number of directions in both NL$_1$ and NL$_2$ subspaces, so it is not a good choice in our setup.} The associated group manifold is $SU(16)$, which has 255 distinct directions. We consider a definition of $k$-locality with $k=4$, which implies that the easy/local directions are given by generators with at most four fermions \cite{Balasubramanian_2020}:
\begin{equation}
T_{\alpha }=\left\{ \psi _{i},\psi _{i}\psi _{j},\psi _{i}\psi _{j}\psi
_{k},\psi _{i}\psi _{j}\psi _{k}\psi _{l}\right\} ,\text{ for }%
i>j>k>l=1,2,\ldots,8,\label{easyd}
\end{equation}%
resulting in 162 easy directions in total. Next, we define the set of five-fermion operators as the least non-local directions on the group manifold (56 of them)
\begin{equation}
T_{\dot{\alpha}_{1}}=\left\{ \psi _{i}\psi _{j}\psi _{k}\psi _{l}\psi
_{m}\right\}, \text{ for }i>j>k>l>m=1,2,...,8,\label{hardd1}
\end{equation}%
with associated cost factor $\mu _{1}$, while the terms with six or more fermions are the hardest/most non-local directions, with cost factor $\mu _{2}$ (37 of them). 


\subsubsection{Three-body Hamiltonian}

First, let us consider the Hamiltonian in \eqref{Hsykchaotic} with three fermions, $H_\textrm{SYK3}$. The corresponding Jacobi equations take the form
\begin{align}
    \frac{d\delta y^{\dot{\alpha}_1}}{ds}&= 
   -it \frac{\mu_2}{1+\mu_1}\mathrm{C}_{\dot{\beta}_2}^{\dot{\alpha}_1}\delta y^{\dot{\beta}_2}, \cr 
     \frac{d\delta y^{\dot{\alpha}_2}}{ds}&= 
   -it  \frac{\mu_1}{1+\mu_2}\mathrm{C}_{\dot{\beta}_1}^{\dot{\alpha}_2}\delta y^{\dot{\beta}_1}
    - it \frac{\mu_2}{1+\mu_2}\mathrm{C}_{\dot{\beta}_2}^{\dot{\alpha}_2}\delta y^{\dot{\beta}_2}\label{h3eq1}.
\end{align}
This system can be reduced to a single second-order differential equation for the component of the most non-local directions:
\begin{equation}
    \frac{d^2 \delta y^{\dot{\alpha}_2}}{ds^2}+
   i t  \frac{\mu_2}{1+\mu_2}\mathrm{C}_{\dot{\beta}_2}^{\dot{\alpha}_2}
    \frac{d\delta y^{\dot{\beta}_2}}{ds}+
   t^2  \frac{\mu_1\mu_2}{(1+\mu_2)(1+\mu_1)}\mathrm{C}_{\dot{\beta}_1}^{\dot{\alpha}_2}
    \mathrm{C}_{\dot{\beta}_2}^{\dot{\beta}_1}\delta y^{\dot{\beta}_2}=0.
\end{equation} 
For $k=4$, the system admits a non-local basis for $\mathrm{NL}_2$, in which the coupling matrices in the last two terms above can be diagonalized simultaneously. In this new basis, the equations decouple and the evolution of the velocity $\delta \tilde{y}^{\dot{\alpha}_2}$ is governed by
\begin{equation}
     \frac{d^2 \delta 
     \tilde{y}^{\dot{\alpha}_2}}{ds^2}+
    i t M_{\dot\alpha_2}
    \frac{d\delta 
    \tilde{y}^{\dot{\alpha}_2}}{ds}+
    t^2\zeta_{\dot\alpha_2}^2 \delta \tilde{y}^{\dot{\alpha}_2}=0,
    \label{neweqy2}
\end{equation}
where
\begin{equation}
M_{\dot{\alpha}_2}=\frac{\mu_2}{1+\mu_2} \lambda^{(\dot{\alpha}_2)} , \qquad 
\zeta^2_{\dot{\alpha}_2} = \frac{\mu_1 \mu_2 }{(1 + \mu_1)(1 + \mu_2)} \rho^{(\dot{\alpha}_2)}.
\end{equation} 
Here, $\lambda^{(\dot{\alpha}_2)}$ and $\rho^{(\dot{\alpha}_2)}$ denote the eigenvalues of $\mathrm{C}^{\dot{\alpha}_2}_{\dot{\beta}_2}$ and $\mathrm{C}_{\dot{\beta}_1}^{\dot{\alpha}_2} \mathrm{C}_{\dot{\beta}_2}^{\dot{\beta}_1}$, respectively.

Instead of solving \eqref{neweqy2} directly, we notice that there is only one case in which $\zeta_{\dot\alpha_2} \ne 0$, which we take to be when, say, $\dot{\alpha}_2 = D$. The appearance of this special direction can be understood from the fermion-parity operator\footnote{We thank the anonymous referee of PRD for pointing out this observation.} $\Gamma=\psi_1\psi_2\cdots\psi_8$, which satisfies
\begin{equation}
    [H_{\mathrm{SYK3}},\Gamma]=2H_{\mathrm{SYK3}}\Gamma .
\end{equation}
For $N=8$, the right-hand side is a linear combination of five-fermion operators and therefore lies entirely in $\mathrm{NL}_1$. Hence, its projection onto $\mathrm{NL}_2$ vanishes, implying $M_D=0$. Moreover, since there exists only one eight-fermion direction, namely $\Gamma$, this is the unique direction in $\mathrm{NL}_2$ that couples to $\mathrm{NL}_1$, explaining why only one $\zeta_{\dot\alpha_2}$ is non-zero. Therefore, we can split the most non-local directions into these two cases, obtaining
\begin{align}
    \delta\tilde y^{\dot\alpha_2}(s)&=\delta\tilde y^{\dot\alpha_2}(0)e^{-itM_{\dot\alpha_2}s}, \qquad \text{for} \quad \dot\alpha_2\ne D, \cr
    \delta\tilde y^{D}(s)&=\delta\tilde y^{D}(0)\cos(t\zeta_D s)-\frac{i}{\zeta_D}\frac{\mu_1}{1+\mu_2}\mathrm{C}^{D}_{\dot\beta_1}\delta y^{\dot\beta_1}(0)\sin(t\zeta_D s).
    \label{jacobisyk3}
\end{align}
The other components involve long expressions and are omitted here. However, the first level of non-locality can be recovered by integrating the first line in \eqref{h3eq1}.


\subsubsection{Four-body Hamiltonian}

In this case, the superoperators $[H_\textrm{SYK4},T_{\dot{\beta}_p}]_{\mathrm{NL}_p}$, with $p=1,2$, are null. Thus, the Jacobi equations in the hard subspaces can be written as
\bea
\frac{d\delta y^{\dot{\alpha}_{1}}}{ds} &=& -it \frac{\mu_2}{1 + \mu_1} \mathrm{C}_{\dot{\beta}_2}^{\dot{\alpha}_1} \delta y^{\dot{\beta}_2}, \cr
\frac{d\delta y^{\dot{\alpha}_{2}}}{ds} &=& -it \frac{\mu_1}{1 + \mu_2} \mathrm{C}_{\dot{\beta}_1}^{\dot{\alpha}_2} \delta y^{\dot{\beta}_1}.
\label{nlj2}
\eea
At this point, we introduce a new basis for the first level of non-locality, $\tilde{T}_{\dot{\beta}_{1}}$, leading to the eigenvalue equation:
\begin{equation}
\big[ H _\textrm{SYK4} ,\big[ H_\textrm{SYK4},\tilde{T}_{\dot{\beta}_1}\big]_{\text{NL}_2} \big]_{\text{NL}_1} = \lambda^{(\dot{\beta}_1)} \tilde{T}_{\dot{\beta}_1}. \label{ssop}
\end{equation}%
The reason for this will become clear later. The strategy for solving the system of differential equations (\ref{nlj2}) is straightforward. Taking the derivative of the first equation and using (\ref{nlj2}), we obtain a second-order equation for the least non-local directions:
\begin{equation}
\frac{d^2 \delta y^{\dot{\alpha}_{1}}}{ds^2} = -t^2 \frac{\mu_1\mu_2}{(1 + \mu_1)(1 + \mu_2)} \mathrm{C}_{\dot{\beta}_2}^{\dot{\alpha}_1} \mathrm{C}_{\dot{\beta}_1}^{\dot{\beta}_2} \delta y^{\dot{\beta}_1}.
\end{equation}%
The new basis introduced in (\ref{ssop}) can be used to decouple the dynamics in the first non-local directions. For the new components $\delta \tilde{y}^{\dot{\alpha}_1}(s)$, we obtain the solutions
\begin{equation}
\delta \tilde{y}^{\dot{\alpha}_1}(s) = \delta \tilde{y}^{\dot{\alpha}_1}(0) \cos(t\zeta_{\dot{\alpha}_1}s)-\frac{i}{\zeta_{\dot{\alpha}_1}}\frac{\mu_2}{1+\mu_1}\mathrm{C}^{\dot{\alpha}_1}_{\dot{\beta}_2}\delta y^{\dot{\beta}_2}(0)\sin(t\zeta_{\dot{\alpha}_1}s),\label{dya1sol}
\end{equation}%
where
\begin{equation}
\zeta_{\dot{\alpha}_1} = \sqrt{\frac{\mu_2 \mu_1 \lambda^{(\dot{\alpha}_1)}}{(1 + \mu_1)(1 + \mu_2)}}.
\end{equation}
With this in hand, we easily obtain the solution for the most non-local subspace
\begin{eqnarray}
\delta y^{\dot{\alpha}_2}(s) &=& \delta y^{\dot{\alpha}_2}(0) - it \frac{\mu_1}{1 + \mu_2}  \mathrm{C}_{\dot{\beta}_1}^{\dot{\alpha}_2} \delta \tilde{y}^{\dot{\beta}_1}(0) \frac{\sin(t\zeta_{\dot{\beta}_1}s)}{t\zeta_{\dot{\beta}_1}}
\cr 
&& \hskip 1cm +t\frac{\mu_1\mu_2}{(1+\mu_1)(1+\mu_2)}\mathrm{C}^{\dot{\alpha}_2}_{\dot{\beta}_1}\mathrm{C}^{\dot{\beta}_1}_{\dot{\beta}_2}\delta y^{\dot{\beta}_2}(0)\frac{\cos(t\zeta_{\dot{\beta}_1}s)-1}{t\zeta^2_{\dot{\beta}_1}}.
\label{nl2syk4}
\end{eqnarray}%
Finally, the local part of the Jacobi field is obtained from the differential equation:
\begin{equation}
\frac{d \delta y^{\alpha}}{ds} = -it\mu_1  \mathrm{C}_{\dot{\beta}_1}^{\alpha} \delta y^{\dot{\beta}_1} -it \mu_2  \mathrm{C}_{\dot{\beta}_2}^{\alpha} \delta y^{\dot{\beta}_2},
\end{equation}
with solution
\bea
\delta y^{\alpha}(s) &=& \delta y^{\alpha}(0) + t \frac{\mu_1\mu_2}{1+\mu_1}  \mathrm{C}_{\dot{\alpha}_1}^{\alpha}\mathrm{C}^{\dot{\alpha}_1}_{\dot{\beta}_2} \delta y^{\dot{\beta}_2}(0) \frac{\cos(t\zeta_{\dot{\beta}_1}s)-1}{t\zeta^2_{\dot{\beta}_1}}\nonumber \\
&& \hskip 1cm  - it\mu_1  \mathrm{C}_{\dot{\beta}_1}^{\alpha} \delta \tilde{y}^{\dot{\beta}_1}(0) \frac{\sin(t\zeta_{\dot{\beta}_1}s)}{t\zeta_{\dot{\beta}_1}}-it\mu_2\mathrm{C}^{\alpha}_{\dot{\alpha}_2}\delta y^{\dot{\alpha}_2}(0)s.
\label{LKYK4}
\eea
Note that the only contribution to the local part coming from \eqref{nl2syk4} is the constant term, since the other terms are proportional to $ \mathrm{C}_{\dot{\beta}_2}^{\alpha}\mathrm{C}_{\dot{\beta}_1}^{\dot{\beta}_2}$, which can be written in terms of the Hamiltonian and the new basis as
\begin{equation}
\mathrm{C}_{\dot{\beta}_2}^{\alpha} \mathrm{C}_{\dot{\beta}_1}^{\dot{\beta}_2} = \frac{1}{2^{N/2}} \mathrm{Tr} \left( T_{\alpha} \big[ H_\textrm{SYK4}, \big[ H_\textrm{SYK4}, \tilde{T}_{\dot{\beta}_1} \big]_{\mathrm{NL}_2} \big] \right).
\end{equation}
It is not hard to see that this term vanishes if \eqref{ssop} is satisfied. 


\subsubsection{Spectral analysis}

\begin{figure}[t]
\centering
\includegraphics[scale=0.55]{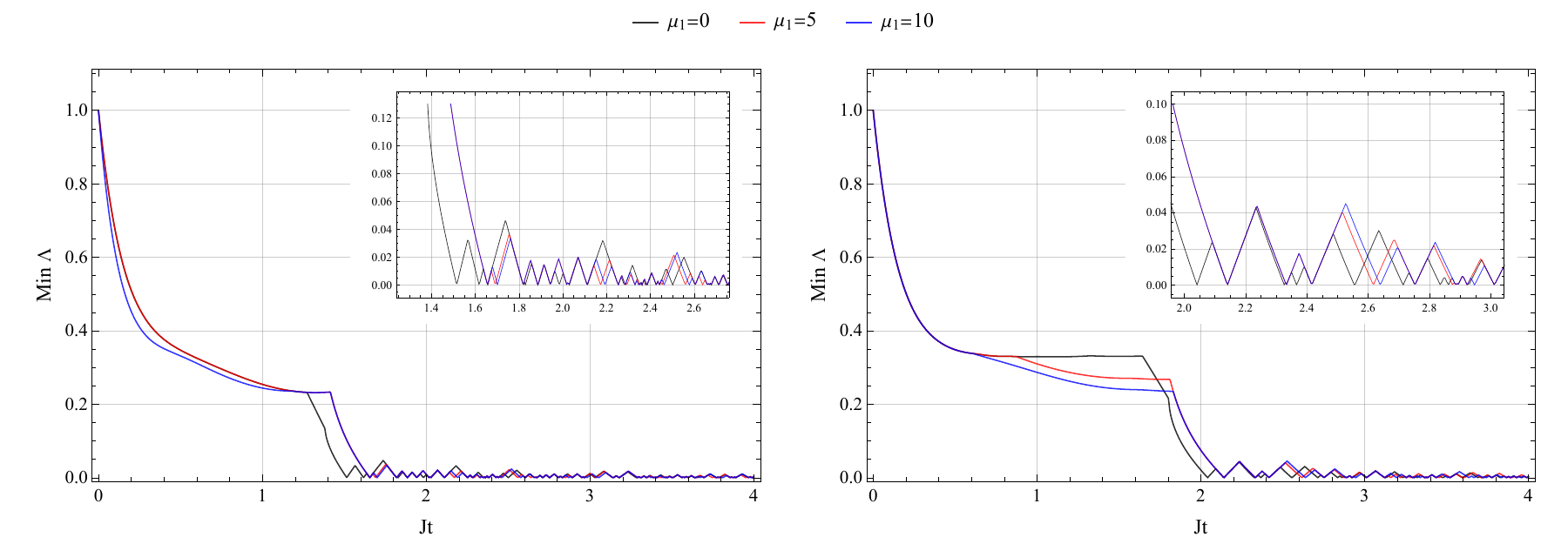}
\caption{Time evolution of the smallest eigenvalue of the Jacobi operator for $N=8$ and $\mu_2=15$, for selected values of the parameter $\mu_1$. The left panel shows the Hamiltonian with three fermions, while the right panel shows the one with four fermions.}\label{chaoticmu1var}
\end{figure}
Here we present the numerical results for the spectrum of the Jacobi operator $\mathbf{Y}_{\mu}[\delta y(0)]$. We follow the same procedure as in section \ref{syk2}, constructing its matrix representation in a convenient basis. For the Hamiltonian $H_{\mathrm{SYK3}}$, we choose the basis $\{T_{\alpha}, T_{\dot\alpha_1}, \tilde{T}_{\dot\alpha_2}\}$, where the bases for $\mathrm{L}$ and $\mathrm{NL}_1$ are given by \eqref{easyd} and \eqref{hardd1}, respectively. The new basis $\tilde T_{\dot\alpha_2}$ for the hardest subspace is chosen to satisfy
\begin{equation}
 \big[ H_\textrm{SYK3},\tilde{T}_{\dot{\alpha}_2}\big]_{\text{NL}_2}
=\lambda^{(\dot\alpha_2)}\tilde{T}_{\dot{\alpha}_2}.
\end{equation}

In contrast, in the four-body case we choose $\{T_{\alpha}, \tilde{T}_{\dot\alpha_1},T_{\dot\alpha_2}\}$, keeping the local basis $T_\alpha$ and the basis $T_{\dot\alpha_2}$, while choosing a new basis for the least non-local subspace satisfying equation \eqref{ssop}. The individual matrix elements are given by
\begin{equation}
(\mathbf{Y}_\mu)_{IJ}=\mathrm{Tr}\left(T^{\dagger}_{I}\mathbf{Y}_\mu[T_J]\right)
\end{equation}
and the explicit form of the Jacobi operator is presented in appendix \ref{FullY}. 
\begin{figure}[t]
\centering
\includegraphics[scale=0.55]{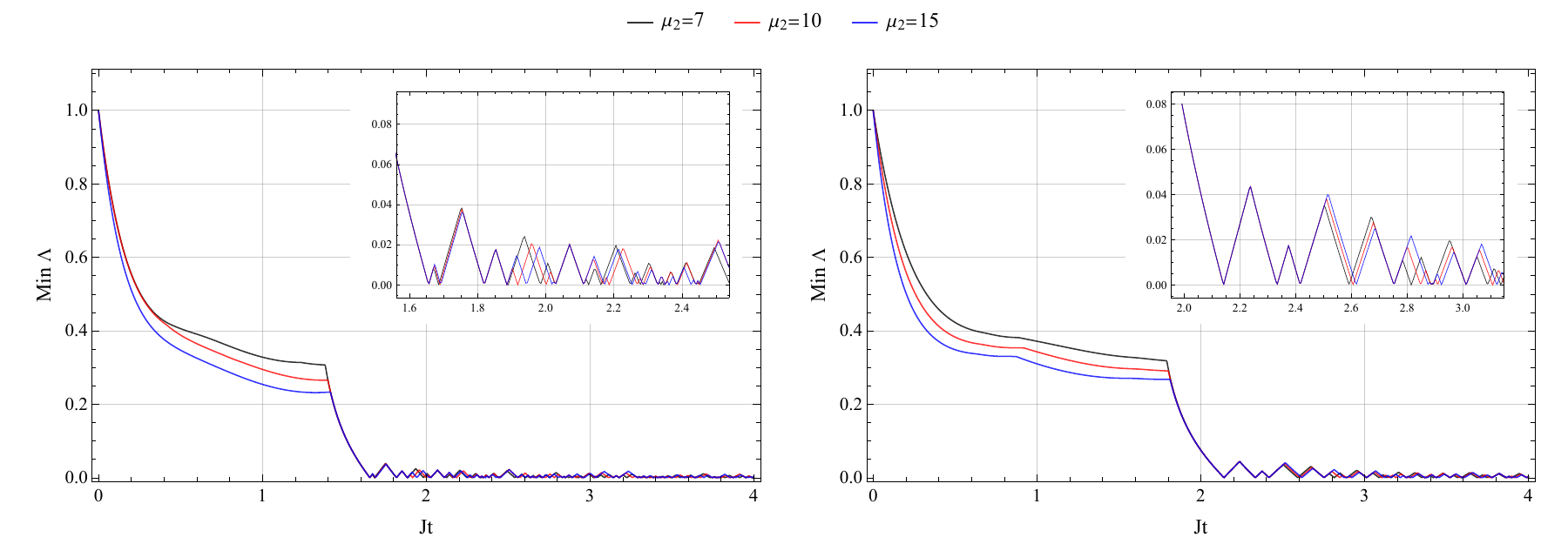}
\caption{Time evolution of the smallest eigenvalue of the Jacobi operator for $N=8$ and $\mu_1=5$, for selected values of the parameter $\mu_2$. The left panel shows the Hamiltonian with three fermions, while the right panel shows the one with four fermions.}\label{chaoticmu2var}
\end{figure}

To determine the location of conjugate times, we compute the time evolution of the smallest Jacobi operator eigenvalue for different values of the cost factors, as shown in figures \ref{chaoticmu1var} and \ref{chaoticmu2var}. In both the three- and four-body Hamiltonians, the conjugate times split into two qualitatively distinct families. The first family is essentially $\mu_i$-independent and remains fixed as the cost factors vary. This family can be interpreted as local conjugate points. The second family depends on the cost factors and is naturally associated with the non-local sectors. These non-local conjugate times are shifted to later times as the corresponding cost factors are increased.
\begin{figure}[t]
    \begin{minipage}[t]{.9\textwidth}
        \centering
        \includegraphics[width=\textwidth]{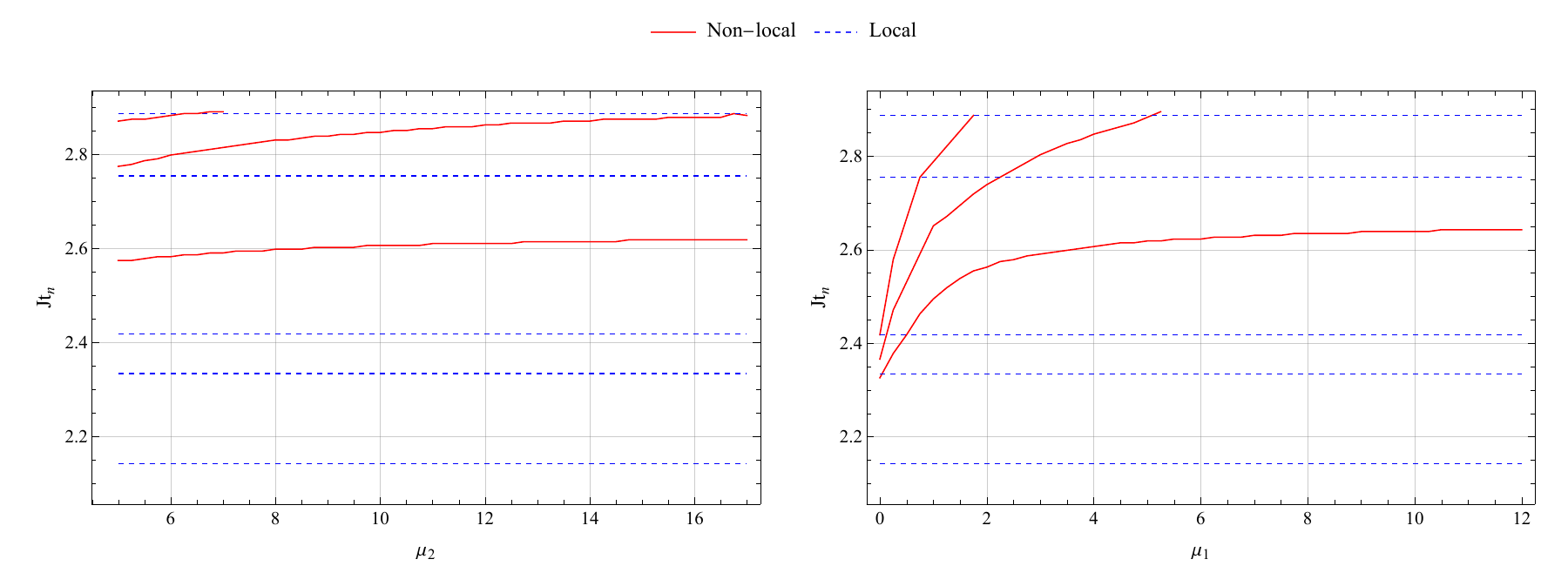}\label{plot1}
    \end{minipage}
    \vfill
      \begin{minipage}[t]{.9\textwidth}
        \centering
        \includegraphics[width=\textwidth]{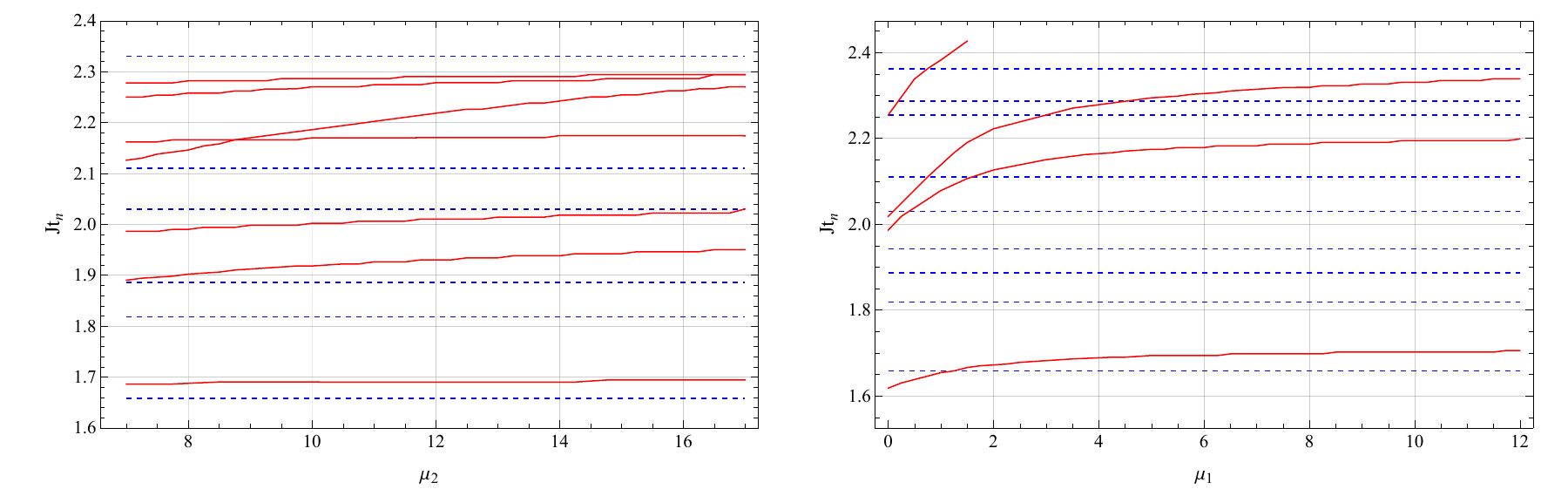}\label{plot3}
    \end{minipage}  
    \caption{Behavior of the first conjugate times, both local and non-local, for the four-body (top) and three-body (bottom) Hamiltonians. The left panels show the variation with $\mu_2$ at fixed $\mu_1=5$, while the right panels show the variation with $\mu_1$ at fixed $\mu_2=15$. Each curve corresponds to a single conjugate time $t_n$ and shows how its occurrence time changes as the corresponding cost factor is varied. The non-local conjugate times are shifted to later times as the cost factors increase, whereas the local conjugate times remain approximately fixed. The three-body case displays a denser and less symmetric pattern, especially under variations of the cost factors.}
    \label{timesvsmus}
\end{figure}

The three-body Hamiltonian exhibits a richer structure than the four-body case. This difference comes from the dynamics in the hardest sector. In the three-body case, the equation for the components $\delta y^{\dot\beta_2}\in\mathrm{NL}_2$ contains an internal contribution governed by the parameter $M_{\dot\alpha_2}$. As a result, varying $\mu_2$ changes not only the coupling between the hard subspaces, $\mathrm{NL}_1$ and $\mathrm{NL}_2$, but also the internal evolution within $\mathrm{NL}_2$. This produces an asymmetry between the variations with respect to $\mu_1$ and $\mu_2$, and leads to a denser pattern of early-time conjugate points than in the four-body Hamiltonian, as shown in figure \ref{timesvsmus}. By contrast, the four-body Hamiltonian displays a more symmetric behavior under variations of $\mu_1$ and $\mu_2$, since the non-local dynamics is mainly controlled by the parameter $\zeta_{\dot\alpha_1}$.


\section{Conclusions}

In this work, we have extended Nielsen’s geometric framework for quantum complexity by introducing a hierarchy of cost factors to distinguish between different classes of non-local operations, yielding a more refined and realistic description of complexity geometry. Within this setup, we have derived the corresponding extensions of the Euler–Arnold and Jacobi equations and analyzed how the presence of multiple penalties modifies both geodesic dynamics and the structure of conjugate points.

A key result of our analysis is that the introduction of multiple cost factors leads to a richer pattern of conjugate points, reflecting the internal structure of the non-local sectors. In particular, the modified Euler-Arnold equations explicitly show that different hard sectors are dynamically coupled through coefficients controlled by the corresponding penalties. The Jacobi equations inherit this structure and become a coupled system for perturbations along the non-local directions. For the case of two hard sectors, we have shown that the system can be treated in Laplace space, where the effect of the most non-local sector on the least non-local one appears as an effective self-energy term $\Sigma^{\ \dot{\alpha}_1}_{\dot{\beta}_1}$ in equation \eqref{defmatrices}. Moreover, we have found that different classes of hard directions generate distinct families of conjugate points, whose locations depend both on the hierarchy of penalties and on the spectral properties of the underlying Hamiltonian. For instance, when the hard sectors do not mix, the result reduces to a simple decoupled generalization of the usual single-penalty case, with the conjugate times associated with a $p$-hard direction being shifted by $(1+\mu_p)$. When different hard sectors mix, however, the Jacobi operator develops a richer zero-mode structure. In this case, the location of conjugate points depends not only on the size of the penalties, but also on the relative hierarchy between them. 

We have illustrated the formalism in two concrete settings: a single-qubit system with two cost factors and SYK-type models in both free and interacting regimes. For the single-qubit case, the Euler--Arnold equations form a non-linear but integrable system, allowing us to obtain exact solutions in terms of two conserved quadratic quantities. These solutions were then used to reconstruct the corresponding geodesics numerically and to determine the complexity for different target Hamiltonians. We found that the effect of the penalty hierarchy depends strongly on the direction of the target evolution. In particular, evolution entirely along the easy direction is insensitive to the cost factors, while targets involving penalized directions exhibit a nontrivial competition between different geodesic sectors and a pronounced dependence on the relative penalties. In the SYK examples, multiple cost factors generate distinct families of conjugate points associated with the different non-local sectors, whose locations are controlled by both the penalty hierarchy and the algebraic structure encoded in the action of the superoperator in \eqref{Coperator}.

A more detailed analysis of the conjugate-time structure in the SYK models reveals that, in the free SYK case, the Jacobi operator becomes block diagonal in a suitable basis, allowing us to separate the spectrum into local and non-local contributions. The local conjugate times remain essentially independent of the cost factors, while the non-local conjugate times are shifted to later times as the corresponding penalties are increased. This provides an important consistency check of the generalized formalism. In the interacting SYK cases, we again find a separation between cost-independent local conjugate times and cost-dependent non-local ones. However, the detailed pattern depends on the structure of the Hamiltonian. For the four-body Hamiltonian, the response to variations of $\mu_1$ and $\mu_2$ is comparatively symmetric, since the non-local dynamics is mainly controlled by the coupling between the two hard sectors. For the three-body Hamiltonian, on the other hand, the hardest sector also exhibits an internal evolution governed by the corresponding block of the Jacobi equations. This produces a denser distribution of early conjugate points and a more asymmetric response to variations of the cost factors.

Recently, conjugate points in the SYK model with $q=4$ and $N=8$ were also analyzed in \cite{Craps:2025kub} using a different penalty structure. Although that work adopts a different organization of the generators in the algebra, splitting the operator space into even and odd sectors, the authors found a qualitatively similar behavior under variations of the relative penalty between the two sectors. In particular, the conjugate points are shifted to larger values of the geodesic parameter as the relative penalty is increased, in qualitative agreement with the behavior found here. From this perspective, our results provide a complementary indication that the location of conjugate points is controlled by the interplay between the algebraic structure of the operator space and the choice of complexity metric.

Overall, our results emphasize that incorporating multiple cost factors enriches the geometric approach to quantum complexity, revealing new dynamical features that are not present in simpler single-penalty models and that are worthy of further study. 


\subsection*{Acknowledgments}

We are grateful to Gian Camilo, Viktor Jahnke, Felipe Soares S\'a and Daniel Teixeira for collaboration during the initial stages of this work, to Barbara Amaral for feedback and to Pierre-Louis Giscard for useful correspondence. MRR thanks CNPq for financial support under grant number 141276/2021-5. DT is supported in part by the {\it Gauge and String Theory (GAST)} initiative of the Istituto Nazionale di Fisica Nucleare (INFN) of Italy, by the PRD 2026 funds at UniMORE, and by the FAPESP {\it Tem\'atico} grants number 2019/21281-4 and 2024/15298-0. 


\appendix

\section{The SYK Jacobi operator: explicit matrix elements}
\label{FullY}

The final expressions for the matrix components $(\mathbf{Y}_\mu)_{IJ} $ of the Jacobi operator in the SYK case discussed in section \ref{sec:SYK} are quite involved. We collect them in this appendix to avoid cluttering the main text. 

In the Hamiltonian eigenstate basis, the matrix elements $\mathrm{C}^{I}_{J}$ read
\begin{equation}
\mathrm{C}^{I}_{J}=\frac{1}{2^{N/2}}\sum_{m,n}\Delta_{mn}c^{(I)*}_{mn}c^{(J)}_{mn},
\end{equation}
and the local part of the Jacobi operator is given, for both $H_\mathrm{SYK3}$ and $H_\mathrm{SYK4}$, by
\begin{equation}
(\mathbf{Y}_\mu^{\mathrm{L}})_{IJ}=\sum_{m,n}c^{(I)*}_{mn}c^{(J)}_{mn}\phi(t\Delta_{mn}), \qquad J\in \mathrm{L}.
\end{equation}


\subsection{Three-body case}

The first non-local submatrix consists of five different terms
\begin{equation}
(\mathbf{Y}_\mu^{\textrm{NL}_1})_{IJ}= Y^{(1)}_{IJ}+Y^{(2)}_{IJ}+Y^{(3)}_{IJ}+Y^{(4)}_{IJ}+Y^{(5)}_{IJ}, \qquad J\in \mathrm{NL}_1,
\end{equation}
with 
\begin{align}
    Y^{(1)}_{IJ}&= i\frac{\mu_{1}}{2^{N/2}\zeta_D(1+\mu_2)} \sum_{m,n,p,q}\Delta_{pq}c_{pq}^{(D)*}c_{mn}^{(I)*} c_{pq}^{(J)}c_{mn}^{(D)}\frac{\zeta_D +\left(i\Delta_{mn}\sin{(t\zeta_D)}-\zeta_D
\cos{(t\zeta_D)}\right)e^{i\Delta_{mn}t}}{t(\Delta^2_{mn}-\zeta_D^{2})}, 
    \nonumber\\
Y^{(2)}_{IJ}&=\sum_{m,n}c^{(I)*}_{mn}c^{(J)}_{mn}
    \phi\left(t\Delta_{mn}\right), 
    \nonumber\\
    Y^{(3)}_{IJ}&= \frac{\mu_1}{2^{N/2}} \sum_{\alpha}\sum_{m,n,p,q}\Delta_{pq}c^{(\alpha)*}_{pq}c^{(I)*}_{mn} c^{(J)}_{pq}c^{(\alpha)}_{mn}\frac{\phi(t\Delta_{mn})-e^{it\Delta_{mn}}}{\Delta_{mn}},
    \nonumber\\
    Y^{(4)}_{IJ}&=-\frac{\mu_1^2\mu_2}{2^{3N/2}\zeta_D^{3}(1+\mu_1)(1+\mu_2)} \sum_{\alpha,\dot\alpha_1}\sum_{m,n,p,q,r,s,x,y}\Delta_{pq}\Delta_{rs}
\Delta_{xy}c^{(\alpha)*}_{pq}c^{(\dot{\alpha}_1)*}_{rs}  c^{(D)*}_{xy}c^{(I)*}_{mn}c^{(J)}_{xy}c^{(D)}_{rs} c^{(\dot{\alpha}_1)}_{pq}c^{(\alpha)}_{mn}
\nonumber\\
    &\hskip 2cm \times 
\left(\zeta_D\frac{\phi\left(t\Delta_{mn}\right)-e^{it\Delta_{mn}}}{\Delta_{mn}}
+\frac{\left[i\Delta_{mn}\sin{(t\zeta_D)}-\zeta_D\cos{(t\zeta_D)}\right]e^{it\Delta_{mn}}+\zeta_D}{it(\Delta_{mn}^{2}-\zeta_D^2)}\right) ,
    \nonumber\\
     Y^{(5)}_{IJ}&=- \frac{\mu_1\mu_2}{\zeta_D^2 2^{N}(1+\mu_1)(1+\mu_2)}\sum_{\dot\alpha_1}\sum_{m,n,p,q,r,s}\Delta_{pq}\Delta_{rs}c^{(\dot{\alpha}_1)*}_{pq}c^{(D)*}_{rs}c^{(I)*}_{mn}c^{(J)}_{rs}c^{(D)}_{pq}c^{(\dot{\alpha}_1)}_{mn}\nonumber\\
    &\hskip 2cm \times \left( \phi(t\Delta_{mn})+\frac{\left[i\zeta_D\sin{(t\zeta_D)}-\Delta_{mn}\cos{(t\zeta_D)}\right]e^{it\Delta_{mn}}+\Delta_{mn}}{it(\Delta_{mn}^{2}-\zeta_D^2)}\right).
\end{align}
The second level of non-locality splits into two parts: when $J\ne D$ and when $J= D$. The first part is composed by a summation of two different terms, namely
\begin{equation}
(\mathbf{Y}_\mu^{\textrm{NL}_2})_{IJ}=Y^{(6)}_{IJ}+Y^{(7)}_{IJ}, \qquad J \in \mathrm{NL}_2, \quad \text{with} \quad J\ne D, 
\end{equation}
where
\begin{align}
    Y^{(6)}_{IJ}&=\frac{\mu_2}{2^{N/2}}\sum_{\alpha}\sum_{m,n,p,q}\Delta_{pq}c^{(\alpha)*}_{pq}c^{(I)*}_{mn}c^{(J)}_{pq}c^{(\alpha)}_{mn}\frac{\phi\left(t(\Delta_{mn}-M_J)\right)-\phi\left(t\Delta_{mn}\right)}{M_J}, 
    \nonumber\\
Y^{(7)}_{IJ}&=\sum_{m,n}c^{(I)*}_{mn}c^{(J)}_{mn}\phi\left(t(\Delta_{mn}-M_J)\right).\label{Yijnl2}
\end{align}
The second part is composed by three other terms:
\begin{equation}
(\mathbf{Y}_\mu^{\textrm{NL}_2})_{IJ }=Y^{(8)}_{ID}+Y^{(9)}_{ID}+Y^{(10)}_{ID}, \qquad J \in \mathrm{NL}_2, \quad \text{with} \quad J= D, 
\end{equation}
where
\begin{align}
    Y^{(8)}_{ID}&=-\frac{\mu_1\mu_2}{2^{N}\zeta_D^2
\left(1+\mu_1\right)}
  \sum_{\alpha,\dot\alpha_1}  \sum_{m,n,p,q,r,s}\Delta_{pq}\Delta_{rs}
c^{(\alpha)*}_{pq}c^{(\dot{\alpha}_1)*}_{rs}c^{(I)*}_{mn}
c^{(D)}_{rs}c^{(\dot{\alpha}_1)}_{pq}c^{(\alpha)}_{mn}\nonumber\\
    &\hskip 3cm \times \left(\phi(t\Delta_{mn})-
    \frac{\Delta_{mn}+e^{it\Delta_{mn}}
    \left[i\zeta_D\sin(t\zeta_D)-\Delta_{mn}
    \cos(t\zeta_D)\right]}{it(\zeta_D^2-\Delta_{mn}^2)}
    \right)
    ,\nonumber \\
    Y^{(9)}_{ID}&=-\frac{i\mu_2}{2^{N/2}\zeta_D (1+\mu_1)}
\sum_{\dot\alpha_1}\sum_{m,n,p,q}\Delta_{pq}c^{(\dot{\alpha}_1)*}_{pq}c^{(D)}_{pq}
c^{(I)*}_{mn}c^{(\dot{\alpha}_1)}_{mn}\cr
& \hskip 3cm \times
\left(\frac{\zeta_D + \left[
    i \Delta_{mn}  \sin (\zeta_D  t)-\zeta_D  \cos (\zeta_D  t)\right]e^{it \Delta_{mn}}}{t \left(\zeta_D ^2-\Delta_{mn} ^2 \right)}\right),\nonumber \\
    Y^{(10)}_{ID}&=i\sum_{m,n} c^{(I)*}_{mn}c^{(D)}_{mn}\frac{\Delta _{mn}+\left[i \zeta_D  \sin (\zeta_D  t)-\Delta _{mn} \cos (\zeta_D  t)\right]e^{i t \Delta _{mn}} }{t \left(\Delta _{mn}^2-\zeta_D^2\right)}.
\end{align}


\subsection{Four-body case}

A similar analysis for the $H_\mathrm{SYK4}$ case leads to this least non-local submatrix
\begin{equation}
(\mathbf{Y}_\mu^{\mathrm{NL}_1})_{IJ}=Y^{(1)}_{IJ}+ Y^{(2)}_{IJ}+Y^{(3)}_{IJ}, \qquad  J\in \mathrm{NL}_1,
\end{equation}
where
\begin{align}
    Y^{(1)}_{IJ}&=\frac{\mu_1}{2^{N/2}}\sum_\alpha \sum_{m,n,p,q}\Delta_{pq}c^{(\alpha)*}_{pq}c^{(I)*}_{mn}c^{(J)}_{pq}c^{(\alpha)}_{mn}\frac{\left[\zeta_{J}\cos(t\zeta_{J})-i\Delta_{mn}\sin(t\zeta_{J})\right]e^{it\Delta_{mn}}-\zeta_{J}}{it\zeta_{J}(\Delta^2_{mn}-\zeta_{J}^2)},
    \cr
Y^{(2)}_{IJ}&=i\sum_{m,n}c^{(I)*}_{mn}c^{(J)}_{mn}\frac{\left[i\zeta_{J}\sin(t\zeta_{J})-\Delta_{mn}\cos(t\zeta_{J})\right]e^{it\Delta_{mn}}+\Delta_{mn}}{t(\Delta^2_{mn}-\zeta_{J}^2)},
    \cr
    Y^{(3)}_{IJ}&=\frac{\mu_1}{2^{N/2}(1+\mu_2)}\sum_{\dot\alpha_2}\sum_{m,n,p,q}\Delta_{pq}c^{(\dot{\alpha}_2)*}_{pq}c^{(I)*}_{mn}c^{(J)}_{pq}c^{(\dot{\alpha}_2)}_{mn}\frac{\left[\zeta_{J}\cos(t\zeta_{J})-i\Delta_{mn}\sin(t\zeta_{J})\right]e^{it\Delta_{mn}}-\zeta_{J}}{it\zeta_{J}(\Delta^2_{mn}-\zeta_{J}^2)}.
    \cr &
\end{align}
On the other hand, the most non-local contributions read
\begin{equation}
(\mathbf{Y}_\mu^{\mathrm{NL}_2})_{IJ}=Y^{(4)}_{IJ}+ Y^{(5)}_{IJ}+Y^{(6)}_{IJ}+Y^{(7)}_{IJ}+Y^{(8)}_{IJ}, \qquad J\in \mathrm{NL}_2,
\end{equation}
with
\begin{align}
     Y^{(4)}_{IJ}&=-\frac{\mu_1\mu_2}{2^{N}(1+\mu_1)}\sum_{\alpha,\dot\alpha_1}\sum_{m,n,p,q,r,s}\Delta_{pq}\Delta_{rs}c^{(\alpha)*}_{pq}c^{(\dot{\alpha}_1)*}_{rs}c^{(I)*}_{mn}c^{(J)}_{rs}c^{(\dot{\alpha}_1)}_{pq}c^{(\alpha)}_{mn}\nonumber\\
    &\hskip 4cm \times\left(\frac{\phi(t\Delta_{mn})}{\zeta^2_{\dot{\alpha}_1}}+
    \frac{\Delta_{mn}+e^{it\Delta_{mn}}
\left[i\zeta_{\dot{\alpha}_1}\sin(t\zeta_{\dot{\alpha}_1})-\Delta_{mn}
\cos(t\zeta_{\dot{\alpha}_1})\right]}{it\zeta^2_{\dot{\alpha}_1}(\Delta_{mn}^2-\zeta_{\dot{\alpha}_1}^2)}
    \right),
    \cr
    Y^{(5)}_{IJ}&=\frac{\mu_2}{2^{N/2}}\sum_{\alpha}\sum_{m,n,p,q}\Delta_{pq}c_{pq}^{(\alpha)*}c^{(I)*}_{mn}c^{(J)}_{pq}c^{(\alpha)}_{mn}\frac{\phi\left(t\Delta_{mn}\right)-e^{it\Delta_{mn}}}{\Delta_{mn}} ,
    \cr
     Y^{(6)}_{IJ}&=\frac{\mu_2}{2^{N/2}(1+\mu_1)} \sum_{\dot\alpha_1}\sum_{m,n,p,q}\Delta_{pq}c^{(\dot{\alpha}_1)*}_{pq}c^{(I)*}_{mn}c^{(J)}_{pq}c^{(\dot{\alpha}_1)}_{mn}\frac{\left[\zeta_{\dot{\alpha}_1}\cos(t\zeta_{\dot{\alpha}_1})-i\Delta_{mn}\sin(t\zeta_{\dot{\alpha}_1})\right]e^{it\Delta_{mn}}-\zeta_{\dot{\alpha}_1}}{it\zeta_{\dot{\alpha}_1}(\Delta^2_{mn}-\zeta_{\dot{\alpha}_1}^2)},
    \cr
Y^{(7)}_{IJ}&=\sum_{m,n}c^{(I)*}_{mn}c^{(J)}_{mn}\phi\left(t\Delta_{mn}\right),
\cr
 Y^{(8)}_{IJ}&=-\frac{\mu_1\mu_2}{2^{N}(1+\mu_1)(1+\mu_2)}\sum_{\dot\beta_1,\dot\alpha_2}\sum_{m,n,p,q,r,s}\Delta_{pq}\Delta_{rs}c^{(\dot{\alpha}_2)*}_{pq}c^{(\dot{\beta}_1)*}_{rs}c^{(I)*}_{mn}c^{(J)}_{rs}c^{(\dot{\beta}_1)}_{pq}c^{(\dot{\alpha}_2)}_{mn}
 \nonumber\\
    &\hskip 4cm \times\left(\frac{\phi(t\Delta_{mn})}{\zeta^2_{\dot{\beta}_1}}+    \frac{\Delta_{mn}+e^{it\Delta_{mn}} \left[i\zeta_{\dot{\beta}_1}\sin(t\zeta_{\dot{\beta}_1})-\Delta_{mn} \cos(t\zeta_{\dot{\beta}_1})\right]}{it\zeta^2_{\dot{\beta}_1}(\Delta_{mn}^2-\zeta_{\dot{\beta}_1}^2)}    \right).\cr &
\end{align}


\bibliographystyle{utphys2}
\bibliography{references}
\end{document}